\documentclass[preprint,groupedaddress,showpacs]{revtex4}

\usepackage{graphicx}
\usepackage{epstopdf}
\usepackage{subfigure}
\usepackage{amssymb}
\usepackage{amsmath}
\usepackage{hyperref}
\usepackage{slashed}
\usepackage[capitalize]{cleveref}

\newcommand {\be}{\begin{equation}}
\newcommand {\ee}{\end{equation}}
\newcommand {\bea}{\begin{eqnarray}}
\newcommand {\eea}{\end{eqnarray}}

\newcommand{\bsig}{\mbox{\boldmath$\sigma$}}
\newcommand{\bgamma}{\mbox{\boldmath$\gamma$}}
\newcommand{\bnabla}{\mbox{\boldmath$\nabla$}}
\newcommand{\bfp}{{\bf p}}

\newcommand{\bfq}{{\bf q}}
\newcommand{\bfk}{{\bf k}}
\newcommand{\bfx}{{\bf x}}
\newcommand{\bfl}{{\bf l}}
\newcommand{\bfS}{{\bf S}}
\newcommand{\bfr}{{\bf r}}
\newcommand{\lqt}{\textquotedblleft}
\newcommand{\la}{\langle}
\newcommand{\ra}{\rangle}

\begin{document}

\preprint{NT@UW-10-16}

\title{Impulse approximation in nuclear pion production reactions: absence of a one-body operator}

\author{Daniel R. Bolton}
\author{Gerald A. Miller}
\affiliation{Department of Physics, University of Washington, Seattle, Washington 98195-1560, USA}

\date{\today}

\begin{abstract}
The impulse approximation of pion production reactions is studied by developing a relativistic formalism, consistent with that used to define the nucleon-nucleon potential.  For plane wave initial states we find that the usual one-body (1B) expression ${\cal O}_\text{1B}$ is replaced by ${\cal O}_\text{2B}=-iK(m_\pi/2){\cal O}_\text{1B}/m_\pi$, where $K(m_\pi/2)$ is the sum of all irreducible contributions to nucleon-nucleon scattering with energy transfer of $m_\pi/2$.  We show that ${\cal O}_\text{2B}\approx{\cal O}_\text{1B}$ for plane wave initial states.  For distorted waves, we find that the usual operator is replaced with a sum of two-body operators that are well approximated by the operator ${\cal O}_\text{2B}$.  Our new formalism solves the (previously ignored) problem of energy transfer forbidding a one-body impulse operator.  Using a purely one pion exchange deuteron, the net result is that the impulse amplitude for $np\to d\pi^0$ at threshold is enhanced by a factor of approximately two.  This amplitude is added to the larger ``rescattering" amplitude and, although experimental data remain in disagreement, the theoretical prediction of the threshold cross section is brought closer to (and in agreement with) the data.
\end{abstract}


\maketitle

%%%%%%%%%%%%%%%%%%% INTRODUCTION %%%%%%%%%%%%%%%%%
\section{\label{sec:intro}Introduction}

It has been known for several decades that the chiral symmetry of the strong nuclear force in the $m_q\rightarrow0$ limit can be exploited to formulate an effective field theory using hadrons as fundamental degrees of freedom rather than quarks and gluons \cite{Weinberg:1978kz, Gasser:1983yg, Bernard:1993nj}.  This theory, generically called chiral perturbation theory (ChPT), is widely used in both the mesonic and the $A=1$ sectors.  Much effort is being put into the application of ChPT to the $A=2$ sector, with success at low energies \cite{Epelbaum:2004fk, Entem:2003ft, Bogner:2003wn}.  The frontier of this program is the pion production threshold, where the relative momentum between colliding nucleons is $p=\sqrt{m_\pi m_N}$.  Pion production is also interesting in its own right as it provides a window into three nucleon forces \cite{Baru:2009fm} and can be used to extract information about charge symmetry breaking \cite{Bolton:2009rq, Filin:2009yh}.

Being an effective theory, ChPT contains an infinite number of interactions organized in terms of importance according to a power counting scheme with an expansion parameter of $m_\pi/\Lambda_\chi$ where $\Lambda_\chi\approx m_N$ is the scale at which the theory ceases to become valid.  For the problem of pion production one finds an additional parameter $\chi\equiv p/m_N=m_\pi/p\approx0.4$.  The fact that this parameter is large provides a significant challenge and a reorganized counting scheme was proposed in Ref. \cite{Cohen:1995cc}.

For a nice review of the history of meson production see Ref. \cite{Hanhart:2003pg}.  The present study considers the specific reaction $NN\to d\pi$ (the two reactions $pp\to d\pi^+$ and $np\to d\pi^0$ are related by isospin symmetry), with the pion in an $s$-wave.  Furthermore, we are focusing on the contribution of a specific diagram, the impulse approximation (IA), also known as ``direct" production, in which the produced pion does not interact at all with the spectator nucleon.  We would like to be clear that pion rescattering, not the IA, is known to be the largest contribution to the total cross section. \cite{Koltun:1965yk}  The $\Delta(1232)$ resonance is also known to contribute significantly to this observable.  Our motivation for the present study is to obtain increased precision in the total cross section calculation and to prepare for future application to other observables to which the IA contributes, such as $p$-wave pion production.

An additional challenge in the calculation of pion production is the presence of strongly interacting initial/final states.  Because NN potentials are only now becoming reliable at such high energies, one typically employs a hybrid calculation in which a kernel is calculated perturbatively from ChPT and then convolved with wave functions calculated from phenomenological potentials.  Recently, this method has come under question for the IA \cite{Bolton:2010qu, Gardestig:2005sn}.  Ideally, one would like to derive the correct method from a relativistic formalism that cleanly separates effects in wave functions from those appearing in the kernel.

\begin{figure}
\centering
\includegraphics[height=1.5in]{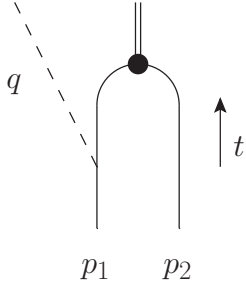}
\caption{\label{fig:iapw}Impulse approximation without initial state interactions.  Solid lines represent nucleons, dashed lines represent pions, and double solid lines represent a deuteron.}
\end{figure}

Consider the IA contribution to $NN\rightarrow d\pi$ in the plane wave (PW) approximation where initial state interactions are neglected (see Fig. \ref{fig:iapw}).  The amplitude for such a process has been estimated to go like ${\cal M}^{IA}\sim\frac{m_\pi}{m_N}\bsig\cdot\mathbf{p}_1\phi(p)\sim  \frac{m_\pi}{m_N}\sqrt{m_Nm_\pi}\phi(p) $, where $\phi(p)$ is the bound state wave function, evaluated in momentum space.
The suppression by $m_\pi^{3/2}$ was noted in Ref. \cite{Cohen:1995cc}, which also included an analysis that a more detailed treatment of the power counting based on including initial and final state interactions introduces a power of $1/m_\pi$ via an energy denominator such that the amplitude varies as $\sqrt{m_\pi}$.  Nevertheless, we see directly an explicit $m_\pi^{3/2}$ times $\phi(\sqrt{m\;m_\pi})$.

In the physical region where $m_\pi=140$ MeV, the wave function falls as a power of momentum greater than unity.  For small values of relative momentum, the deuteron wave function also falls more rapidly than an inverse power of its argument.  If one takes $m_\pi$ to be small, the deuteron remains weakly bound \cite{Beane:2002vs, Bulgac:1997ji} and therefore its momentum wave function will also fall rapidly in the chiral limit.  Thus the power counting can only be considered a very rough estimate.  If we follow \cite{Cohen:1995cc}, the impulse term is a leading order term, but the deuteron wave function is quite small for physical values of $p$ and there is also a substantial cancellation between the deuteron $s$- and $d$-states.  Thus this term's contribution to the cross section \cite{Koltun:1965yk} is small and there is a contradiction between power counting expectations and realistic calculations.

This contradiction was also discussed at length in Ref. \cite{Bolton:2010qu} where the authors introduced \lqt wave function corrections" as a possible solution.  This proposal included one-pion exchange (OPE) with an energy transfer of $m_\pi/2$ in the impulse kernel, but then subtracted off a similar diagram with static OPE in order to prevent double counting.  The result depended strongly on the treatment of the intermediate off-shell nucleon propagator and no definitive conclusion was reached.  This present work is intended to settle the debate regarding the inclusion of OPE in the impulse approximation.  We demonstrate, by starting from a consistent relativistic formalism, that non-static OPE is to be included with no subtraction necessary; the impulse amplitude that should be used is given in Eq. (\ref{eq:mdwfinal}).  Furthermore, we show that the traditional approach of using a one-body kernel is correct only in the absence of initial state interactions.

In Sec. \ref{sec:bs} we review aspects of the Bethe-Salpeter (BS) formalism for the two-nucleon problem.  Section \ref{sec:nnpiamp} presents the $N\to N\pi$ operator and Sec. \ref{sec:pw} shows that for plane wave $NN\to d\pi$, the traditional impulse approximation is approximately valid.  Next, Sec. \ref{sec:distortions} considers the full distorted-wave amplitude by calculating the corresponding loop diagram, including the effects of the non-zero time components of the momenta of the exchanged mesons.  In this section, we are able to interpret the distorted-wave amplitude as a sum of two-body operators.  We demonstrate the new formalism by explicitly evaluating $s$-wave $NN\rightarrow d\pi$ amplitudes at threshold.  To aid the flow of the arguments, approximations made in this section are verified to be sub-leading in \cref{sec:isi,sec:fsi,sec:sigma}.  A comparison with experimental cross section data is made in Sec. \ref{sec:discussion}, where we also discuss implications and future directions.

%%%%%%%%%%%%%%%%% BETHE-SALPETER BASICS %%%%%%%%%%%%%
\section{\label{sec:bs}Bethe-Salpeter Basics}

Recall the definition of the nucleon-nucleon potential from the Bethe-Salpeter formalism. We follow the approach of Partovi and Lomon \cite{Partovi:1969wd} and also consider the relationship between the Bethe-Salpeter wave function and the usual equal time wave function as recently discussed in Ref. \cite{Miller:2009fc}.

Partovi and Lomon write the Bethe-Salpeter equation for the nucleon-nucleon scattering amplitude $\cal M$ as 
\be
{\cal M}=K +K G {\cal M},\label{eq:bs}
\ee
where $K$ is the sum of all irreducible diagrams.  The quantities ${\cal M}$ and $K$ depend on the total four-momentum $P_\text{tot}$ and the relative four-momentum  $k$.  The two individual momenta are $p_{1,2}=P_\text{tot}/2\pm k$ and $G$ is the product of two Feynman propagators:
\be
G =\left(\frac{i}{\slashed{p}_1 -m_N +i\epsilon}\right)_1 \; \left(\frac{i}{\slashed{p}_2 -m_N +i\epsilon}\right)_2=G_1G_2,
\ee
where $m_N$ is the nucleon mass. The quantities ${\cal M}$ and $K$ differ from those of \cite{Partovi:1969wd} by a factor of $-i/(2\pi)$.  Partovi and Lomon replace the relativistic $G$ by the Lippmann-Schwinger propagator $g$ for two particles.  For scalar particles, $g$ is obtained from $G$ by integrating over the zero'th (energy)  component of one of the two particles \cite{Miller:2009fc}. For fermions, one must also project onto the positive energy sub-space of both particles.  This is accomplished in the center of mass frame by taking \cite{Partovi:1969wd}
\be
g(k|P_\text{tot})=2\pi i\frac{[\gamma^0 E(\bfk)-\bgamma\cdot\bfk +m_N]_1[\gamma^0 E(\bfk)+\bgamma\cdot\bfk +m_N]_2}{E(\bfk)(P_\text{tot}^2-4m_N^2-4\bfk^2+i\epsilon)}\delta(k^0) ,\label{eq:g}
\ee
where $E(\bfk )\equiv \sqrt{\bfk^2+m_N^2}$.  Note that $g$ contains the important two-nucleon unitary cut.  The non-relativistic potential $U$ is defined so as to reproduce the correct on-shell NN scattering amplitude $\cal M$ using the Lippmann-Schwinger (LS) equation
\bea {\cal M}=U+Ug {\cal M}.\label{eq:ls}\eea 
The quantity $U$ is obtained by equating the ${\cal M}$ of Eq. (\ref{eq:bs}) with that of Eq. (\ref{eq:ls}) to find \cite{Partovi:1969wd}
\be
U=K+K(G-g)U.\label{eq:udef}
\ee
In solving Eq. (\ref{eq:ls}) for the on-energy shell scattering amplitude, $U$ never changes the value of the relative energy $k^0$ away from 0.  Equations (\ref{eq:ls}) and (\ref{eq:udef}) are consistent with  Weinberg power counting in which one calculates the potential using chiral perturbation theory and then solves the LS equation to all orders.  The term $G-g$ may be thought of a purely relativistic effect arising from off-shell (short-lived) intermediate nucleons, and in the present context a perturbative effect.

\begin{figure}
\centering
\includegraphics[height=1.25in]{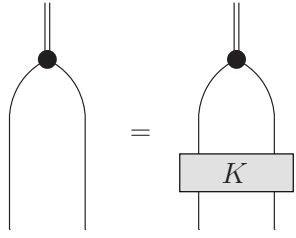}
\caption{\label{fig:dBS}Bethe-Salpeter equation near the deuteron pole.}
\end{figure}

\begin{figure}
\begin{center}
\subfigure[\ $NN$ scattering]{\label{fig:scattBS}\includegraphics[height=1in]{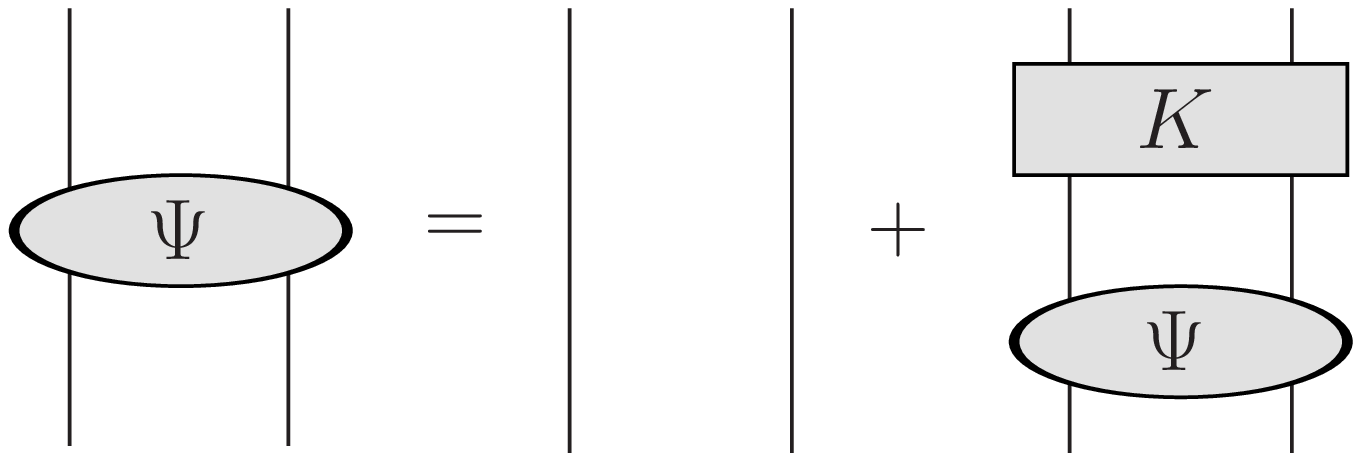}}
\hspace{.15\linewidth}
\subfigure[\ Deuteron]{\label{fig:dwfn}\includegraphics[height=1in]{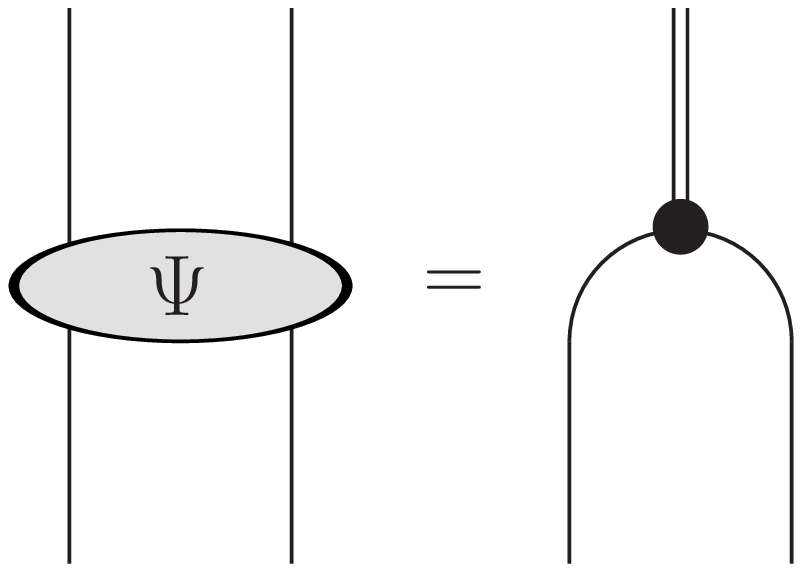}}
\end{center}
\caption{\label{fig:BSwfns}Bethe-Salpeter wave functions.}
\end{figure}

Consider the deuteron wave function in the final state of a pion production reaction. For $P^2$ near the pole position, the second term of Eq. (\ref{eq:bs}) dominates and we replace the scattering amplitude with the vertex function $\Gamma$:  ${\cal M}\rightarrow \Gamma$, and
\be
\Gamma=KG\Gamma.
\ee
This equation is shown pictorially in Fig. \ref{fig:dBS}.
The Bethe-Salpeter wave function $\Psi$ is defined as $G\Gamma$ so that 
\be
\Psi=G\Gamma=GK\Psi.\label{eq:bsb}
\ee 
The wave functions of the scattering state and the deuteron are shown in Fig. \ref{fig:BSwfns}.
If one uses Eq. (\ref{eq:ls}), the bound-state wave function $\phi$ is obtained by solving the equation
\be
\phi=gU\phi=gUgU\phi.
\ee
The second equation shows that $U$ also is evaluated at vanishing values of time components of the relative momenta. We will treat the amplitudes $\Psi,\phi$ and the state vectors $|\Psi\rangle,|\phi\rangle$ (either bras or kets) as interchangeable. 

The next step is to relate $\Psi$ with $\phi$, which can be thought of as the usual bound-state wave function.
 This is most easily accomplished by using the projection operator $P$ on the product space of two positive-energy on-mass-shell nucleons.  We then have
\be
PG=GP\equiv G_P=g,
\ee
with the last step resulting from the explicit appearance of two positive-energy projection operators for on-mass-shell nucleons in Eq. (\ref{eq:g}).  We define $Q=I-P$ and use the notation $\Psi_{P}\equiv {P}\Psi,\Psi_{Q}\equiv {Q}\Psi$ and $PKP\equiv K_{PP},PKQ\equiv K_{PQ}\;,{etc}.$  The $Q$-space includes all terms with one or both nucleons off the mass-shell.  The amplitude $\Psi_P$ contains the ordinary nucleonic degrees of freedom so  one expects that it corresponds to $\phi$. This is now shown explicitly.  Use $I=P+Q$ in Eq. (\ref{eq:bsb}) and multiply by $P$ and then also by $Q$ to obtain the coupled-channel version of the relativistic bound state equation:
\bea
\Psi_P&=&G_PK_{PP}\Psi_P+G_PK_{PQ}\Psi_Q\label{eq:psip}
\\
\Psi_Q&=&G_QK_{QP}\Psi_P+G_QK_{QQ}\Psi_Q.\label{eq:psiq}
\eea
Solving Eq. (\ref{eq:psiq}) for $\Psi_Q$ and using the result in Eq. (\ref{eq:psip}) gives
\bea
\Psi_Q&=&[1-G_QK_{QQ}]^{-1}G_QK_{QP}\Psi_P
\\
\Psi_P&=&G_P\left(K_{PP} +K_{PQ}[G_Q^{-1}-K_{QQ}]^{-1}K_{QP}\right)\Psi_P,\label{eq:peq}
\eea
but one can multiply Eq. (\ref{eq:udef}) by $P\cdots P$ \textit{etc}. to obtain the result
\be
U_{PP} =K_{PP} +K_{PQ}[G_Q^{-1}-K_{QQ}]^{-1}K_{QP},
\ee
thus Eq. (\ref{eq:peq}) can be re-expressed as
\be
\Psi_P=G_PU_{PP}\Psi_P=gU\Psi_P.
\ee
This last equation is identical to Eq. (\ref{eq:bsb}).  Thus we have the result that 
\be
\Psi_P=\phi.
\ee
$\Psi_P$ is not the complete wave function, but we expect that $\Psi_Q$ is a perturbative correction because the deuteron is basically a non-relativistic system.

%%%%%%%%%%%%%%%%%%% NNdpi %%%%%%%%%%%%%%%%%%%%%%%%
\section{\label{sec:nnpiamp}The $N\rightarrow N\pi$ amplitude}

We now turn to the application of the Bethe-Salpeter formalism to the problem of threshold pion production.  First, we remind the reader of the one-body pion production operator in baryon chiral perturbation theory (BChPT) \cite{Jenkins:1990jv}.  For a modern review, see Ref. \cite{Scherer:2002tk}.  In this theory, the nucleon field is split into its heavy ($H_v$) and light ($N_v$) components,
\bea
\Psi(x)&=&e^{-im_Nv\cdot x}\left(N_v(x)+H_v(x)\right)\nonumber
\\
N_v(x)&=&e^{im_Nv\cdot x}P_+\Psi(x)\nonumber
\\
H_v(x)&=&e^{im_Nv\cdot x}P_-\Psi(x)
\eea
where $P_\pm=(1\pm\slashed{v})/2$ and $v$ is the velocity vector satisfying $v^2=1$ and chosen in this work to be $v=(1,{\bf 0})$.  The heavy component is integrated out of the path integral and the resulting free equation of motion for the light component has a solution,
\be
N(x)=\sqrt{E+m_N}\begin{pmatrix}\chi\\0\end{pmatrix}e^{-i(E-m_N)t+i\bfp\cdot\bfx},
\ee
where $E=\sqrt{\bfp\,^2+m_N^2}$ and $\chi$ is a two-component Pauli spinor.  In Appendix \ref{sec:nnpi} we show that the leading order (LO) Feynman rule for the $s$-wave $N\rightarrow N\pi$ amplitude vanishes at threshold and that the next-to-leading order (NLO) rule is
\be
\mathcal{O}_\pi=-i\frac{m_\pi}{2m_N}\frac{g_A}{2f_\pi}\gamma^5\gamma_i\gamma^0(\overrightarrow{\bnabla}-\overleftarrow{\bnabla})_i\tau_a,\label{eq:vertex}
\ee
where the derivatives act on the nucleon wave functions.

%%%%%%%%%%%%%%%%% PLANE WAVE %%%%%%%%%%%%%%%%%%%
\section{\label{sec:pw}The $NN\to d\pi$ reaction: plane wave initial states}

Traditionally, the impulse approximation to pion production is calculated by using the operator of Eq. (\ref{eq:vertex}) as the irreducible kernel to be evaluated between non-relativistic nucleon-nucleon wave functions for the initial and final states.  Between two-component nucleon spinors, $\gamma^5\gamma_i\gamma^0\rightarrow\sigma_i$, so
\be
\mathcal{M}^\text{PW}_\text{1B}=\la\phi|\left[-i\frac{m_\pi}{2m_N}\frac{g_A}{2f_\pi}\bsig\cdot\left(\overrightarrow{\bnabla}-\overleftarrow{\bnabla}\right)\tau_{a}\right]|p_1,p_2\ra,\label{eq:Mapprox}
\ee
where the superscript on ${\cal M}$ indicates that we have neglected initial state interactions.  Next, we show that Eq. (\ref{eq:Mapprox}) is only an approximation to the full impulse amplitude derived from the relativistic Bethe-Salpeter formalism.  We will see that this approximation is only valid in the absence of initial state interactions.

For the case of plane waves in both the initial and final states, a one-body operator is forbidden by energy-momentum conservation,
\be
\la p_3,p_4|{\cal O}_\pi|p_1,p_2\ra=0,
\ee
with all the $p_i$ on mass shell.  The correct formalism must be able to explain the required energy transfer.  Our primary thesis is that the diagram of Fig. \ref{fig:iapw} must be obtained from the Feynman rules as
\be
{\cal M}^\text{PW}_\text{2B}=\la\Gamma|G_1{\cal O}_\pi|p_1,p_2\ra=\left\la\Psi\left|K(m_\pi/2)\,G_1{\cal O}_\pi\right|p_1,p_2\right\ra,\label{eq:mpw}
\ee
where $G_1$ is the Feynman propagator of the intermediate off-shell nucleon and $K(m_\pi/2)$ is the sum of all irreducible diagrams with energy transfer of $m_\pi/2$.  The second equality of Eq. (\ref{eq:mpw}) results from the relation between $\Gamma$ and $\Psi$ in Eq. (\ref{eq:bsb}).  This manipulation is necessary because $\la\phi|$ will be used for evaluation instead of $\la\Psi|$, meaning that the relative energy must remain zero in the final state.  Thus the full kernel for pion production via the impulse approximation is $KG_1{\cal O}_\pi$ rather than just ${\cal O}_\pi$.  Because $KG_1{\cal O}_\pi$ is a two-body operator, the momentum mismatch which suppresses the IA in the traditional treatment is removed.  

There are two points to emphasize here.  Firstly, this treatment is not equivalent to the heavy meson exchange operators of Refs. \cite{Horowitz:1993sh, Lee:1993xh} which are intended to account for the relativistic initial and final state interactions not present in phenomenological potentials.  Secondly, although the assertion of Eq. (\ref{eq:mpw}) greatly changes the way impulse pion production is calculated, one should not perform the same manipulations for the similar impulse approximation to photo-disintegration.  The reason for this is simply that near threshold the nucleon remains essentially on-shell and the diagram is therefore clearly reducible.

Next, we use $\Psi=\Psi_P+\Psi_Q=\phi+\Psi_Q$ and focus on the $\phi$ term; the other term contains non-nucleonic physics and may be treated as a correction.  Thus the impulse approximation is given by
\be
{\cal M}^\text{PW}_\text{2B}\approx\la\phi|K(m_\pi/2)\,G_1{\cal O}_\pi|p_1p_2\ra.\label{eq:Mia}
\ee

Consider the spacetime structure of the product, $G_1{\cal O}_\pi$.  The relativistic propagator $G_1$ is decomposed into three terms: $1$, $\gamma^0$, and $\gamma^i$.  Between two-component nucleon spinors
\bea
\gamma^5\gamma_i\gamma^0&\rightarrow&\sigma_i\nonumber
\\
\gamma^0\gamma^5\gamma_i\gamma^0&\rightarrow&\sigma_i\nonumber
\\
\gamma^i\gamma^5\gamma_j\gamma^0&\rightarrow&0,\label{eq:nrelred}
\eea
and so we can make the replacement
\bea
G_1{\cal O}_\pi=i\frac{\slashed{p}_1-\slashed{q}+m_N}{(p_1-q)^2-m_N^2+i\epsilon}{\cal O}_\pi&\rightarrow&i\frac{E(\bfp_1)-m_\pi+m_N}{-2E(\bfp_1)m_\pi+m_\pi^2+i\epsilon}{\cal O}_\pi\nonumber
\\
&=&\frac{i}{-m_\pi}\left(1-\frac{m_\pi}{4m_N}\right){\cal O}_\pi,
\eea
where in the second line we have used that $E(\bfp_1)=m_N+m_\pi/2$ at threshold.  Note that this propagator agrees with that obtained from the Feynman rules for BChPT at LO.

\begin{figure}
\centering
\includegraphics[height=1.5in]{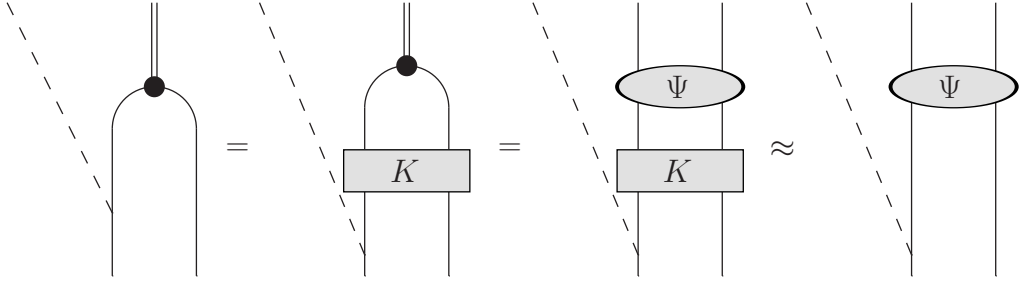}
\caption{\label{fig:iapwBS}Bethe-Salpeter formalism applied to pion production for plane wave initial states.}
\end{figure}

In order to make connection with the traditional Eq. (\ref{eq:vertex}), we use the approximations $K\approx U$ [corrections are ${\cal O}(g-G)$] and $G_1\approx-i/m_\pi$ [corrections are ${\cal O}(m_\pi/m_N)$].  Putting these substitutions into Eq. (\ref{eq:Mia}), 
\be
{\cal M}^\text{PW}_\text{2B}\approx\la\phi|\left[-\frac{iU\left(\frac{m_\pi}{2}\right)}{m_\pi}{\cal O}_\pi\right]|p_1p_2\ra.\label{eq:mpwfinal}
\ee
The quantity $U$ is related to the potential energy by $U=-iV$.  Ignoring the fact that $U$ should be evaluated for non-zero energy transfer, we use the equal-time Schr\"{o}dinger equation to replace $V\rightarrow-E_d-p^2/m_N$ and then neglect the binding energy to find ${\cal M}^\text{PW}_\text{2B}\approx{\cal M}^\text{PW}_\text{1B}$.  This means that for a PW initial state, the traditional impulse approximation should be roughly adequate.  This is borne out in the actual calculation of the reduced matrix elements for Eqs. (\ref{eq:Mapprox}) and (\ref{eq:mpwfinal}),
\bea
A_\text{1B}^\text{PW}&=&-24.0
\\
A_\text{2B}^\text{PW}&=&-25.6,
\eea
where we have used Ref. \cite{Bolton:2009rq}'s definition of the reduced matrix element (we suppress the subscript on Ref. \cite{Bolton:2009rq}'s $A_0$ for clarity) and used the same static phenomenological potential for $V$ (here, Argonne v18 \cite{Wiringa:1994wb}) that is used to calculate the wave functions.  See Fig. \ref{fig:iapwBS} for a pictorial description of this section.

\begin{figure}
\centering
\includegraphics[height=1.5in]{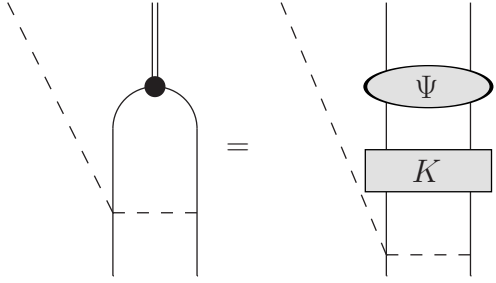}
\caption{\label{fig:rsBS}Use of the Bethe-Salpeter equation in the rescattering amplitude}
\end{figure}

It is important to note that the Bethe-Salpeter equation can also be used for the pion rescattering diagram as shown in Fig. \ref{fig:rsBS}.
In fact, the diagram on the right in Fig. \ref{fig:rsBS} has played an important role in the development of pion production.  The authors of Ref. \cite{Lensky:2005jc} showed that this diagram (with $K$ approximated by OPE) becomes irreducible when the energy dependence of the $NN\pi\pi$ vertex is used to cancel one of the intermediate nucleon propagators.  This discovery resolved a problem arising from calculation of NLO loops.

In the next section, we will show that for distorted wave initial states, Eq. (\ref{eq:mpwfinal}) is replaced by 
\be
{\cal M}^\text{DW}_\text{2B}\approx{}_f\la\phi|\left[-\frac{iU\left(\frac{m_\pi}{2}\right)}{m_\pi}{\cal O}_\pi+{\cal O}_\pi\frac{iU\left(\frac{m_\pi}{2}\right)}{m_\pi}\right]|\phi\ra_i.\label{eq:mdwfinal}
\ee
where the first term contributes at leading order in the theory and the second term at next-to-leading order.

%%%%%%%%%%%%%%%%% DISTORTIONS %%%%%%%%%%%%%%%%%%%
\section{\label{sec:distortions}The $NN\to d\pi$ reaction: distorted wave initial states}
\subsection{\label{sec:dwdef}Definition of distorted wave operator}

There is no reason to expect the result ${\cal M}^\text{PW}_\text{2B}\approx{\cal M}^\text{PW}_\text{1B}$ to carry over for a distorted wave (DW) initial state where $\bfp^2=m_\pi m_N$ no longer holds.
Indeed, we will show that the traditional expression for the impulse approximation does not hold for DW amplitudes.

The fully-relativistic initial-state wave function is denoted $|\Psi\ra_i$,
\be
|\Psi\ra_i=|p_1,p_2\ra+GK|\Psi\ra_i,
\ee
where the first term is exactly the initial state used in the definition of ${\cal M}^\text{PW}$ of Eqs. (\ref{eq:Mapprox}) and (\ref{eq:mpw}).  The complete DW impulse operator is defined as,
\be
{\cal M}^\text{DW}={\cal M}^\text{PW}+{\cal M}^\text{ISI}.\label{eq:mdw}
\ee
The second term includes the production operator $KG_1{\cal O}_\pi$ from Eq. (\ref{eq:mpw}) along with initial state interactions,
\bea
{\cal M}^\text{ISI}_\text{2B}&=&{}_f\la\Psi|KG_1{\cal O}_\pi GK|\Psi\ra_i
\\
&\approx&{}_f\la\phi|KG_1{\cal O}_\pi GK|\phi\ra_i,\label{eq:misi}
\eea
where in the second line we have once again used $\Psi=\phi+\Psi_Q$ and neglected the $Q$-space.

\begin{figure}
\centering
\includegraphics[height=3in]{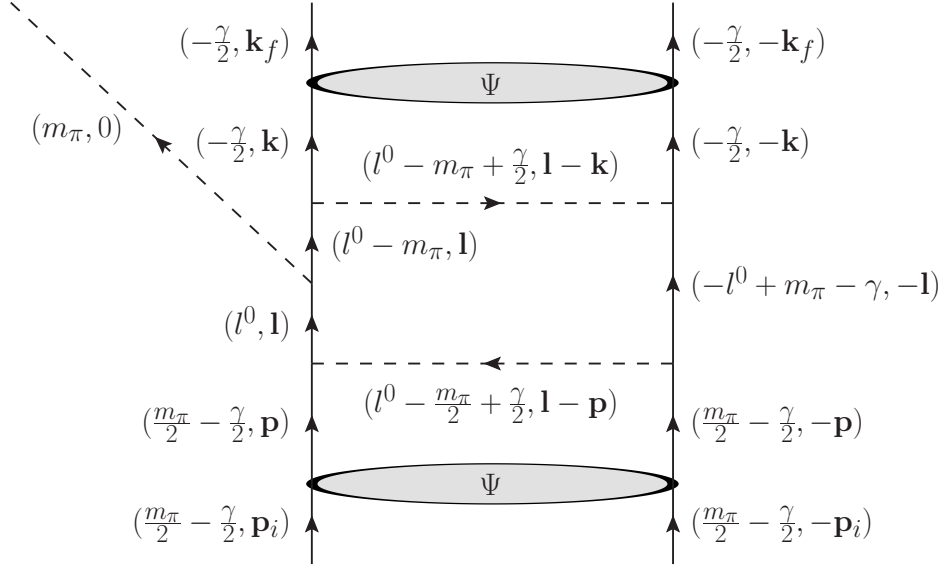}
\caption{Impulse approximation using distorted waves.  Solid lines represent nucleons, dashed lines represent pions, and ovals represent wave functions.}
\label{fig:loop}
\end{figure}

As noted by Ref. \cite{Hanhart}, the kernel of Eq. (\ref{eq:misi}) is a loop integral which is shown in Fig. \ref{fig:loop} with $K$ being approximated by OPE.
Note that four-momenta are conserved at every vertex.  One pion exchange is the first contribution to $K$ in ChPT besides a short range operator which is irrelevant for the $s$-wave $NN\to d\pi$ amplitude (see \cref{sec:lagrangian}).  Nevertheless, one must excercise caution due to the large expansion parameter of pion production.  To this end, we employ the deuteron of Ref. \cite{Friar:1984wi} which is calculated from a purely-OPE potential with suitable form factors.  As discussed in \cref{sec:deuteron}, this deuteron wave function is quite accurate and increases the rescattering amplitude by only 3\% over a phenomenological deuteron.  Having then employed this deuteron wave function in the calculation of the traditional DW impulse approximation, we will be able to avoid any complications from higher order parts of the potential in our subsequent investigation of the two-body operator of Eq. (\ref{eq:misi}).  In other words, although the full potential must be present in an exact calculation, we expect to gain insight into the correct formalism by using an OPE-only deuteron.  We continue to use the phenomenological potentials for the initial state.  To verify that the use of $K=\text{OPE}$ in the initial state does not spoil our results too much, \cref{sec:sigma} examines heavy meson exchange in the initial state.  As will be discussed, this effect is parametrically suppressed.

Note that the relative momenta of the nucleons before and after the loop ($\bfp$ and $\bfk$) are external momenta to the loop integral over $l=(l^0,\bfl)$, but are eventually integrated over in a momentum-space evaluation.  Let us focus solely on the energy part of the loop integral and ignore the vertex factors and overall constants.  We define the integral $I$,
\bea
I&=&i^5\int\frac{dl^0}{2\pi}\frac{1}{l^0-E+i\epsilon}\frac{1}{l^0-m_\pi-E+i\epsilon}\frac{1}{-l^0+m_\pi-\gamma-E+i\epsilon}\frac{1}{l^0-m_\pi/2-\gamma/2+\omega_i-i\epsilon}\nonumber
\\
&&\times\frac{1}{l^0-m_\pi/2-\gamma/2-\omega_i+i\epsilon}\frac{1}{l^0-m_\pi+\gamma/2+\omega_f-i\epsilon}\frac{1}{l^0-m_\pi+\gamma/2-\omega_f+i\epsilon},\label{eq:loop}
\eea
where $\omega_i^2=(\bfl-\bfp)^2+m_\pi^2$ is the on-shell energy of the initial-state pion, $\omega_f^2=(\bfl-\bfk)^2+m_\pi^2$ is the on-shell energy of the final state pion, and $E=\bfl^2/2m_N$ is the kinetic energy of a single intermediate nucleon.  Note that $\bfp_i^2\approx m_\pi m_N-\gamma m_N$ and $\bfk_f^2\approx-\gamma m_N$.

It is straightforward to show that if the energy components of the exchanged pions in the above loop are set to zero (violating conservation of four-momentum), one obtains the traditional impulse approximation.  In this case, the pion energy denominators are pulled out of the integral which is then evaluated by closing the contour in the lower half plane,
\be
I_\text{1B}=\left[\frac{1}{-\omega_f^2}\frac{1}{-\gamma-\bfl^2/m_N}\right]\left[\frac{1}{m_\pi-\gamma-\bfl^2/m_N}\frac{1}{-\omega_i^2}\right].
\ee
The quantity in the first set of brackets can be recognized as the product of OPE with the final state wave function while the second set is the product of the initial state wave function with OPE.  This is precisely the operator that the traditional evaluation includes.

%%%%%%%%%%%%%%%%%%%%%%%%%%%%%%%%%%%%%%%%
\subsection{\label{sec:topt}Reduction to time ordered perturbation theory (TOPT)}
\begin{figure}
\centering
\includegraphics[height=1.25in]{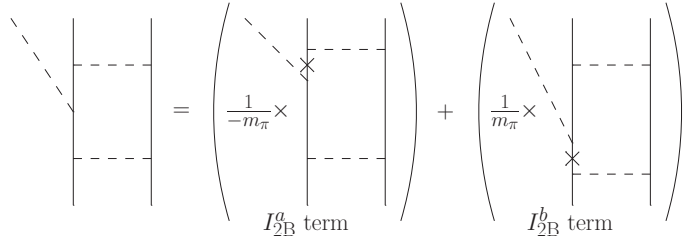}
\caption{Definition of the two terms in Eq. (\ref{eq:split2}).  Crosses represent the propagators which are absent due to the partial fractions decomposition.}
\label{fig:split}
\end{figure}

Our goal is to evaluate the integral in Eq. (\ref{eq:loop}), showing that it is a sum of TOPT terms which can be combined to obtain Eq. (\ref{eq:mdwfinal}).  To begin, we rewrite the first two factors as a sum,
\be
\frac{1}{l^0-E+i\epsilon}\frac{1}{l^0-m_\pi-E+i\epsilon}=\frac{1}{-m_\pi}\left(\frac{1}{l^0-E+i\epsilon}-\frac{1}{l^0-m_\pi-E+i\epsilon}\right).\label{eq:split}
\ee
This is the key to our method because after making this split, we see two terms which each have the propagator structure of a rescattering box loop.  Consider the first term in Eq. (\ref{eq:split}); this loop integral looks like a two-body operator multiplied by $\frac{1}{-m_\pi}$ and augmented with an initial-state interaction.  The second term looks like the same with final-state interaction.  We define these two integrals to be $I_\text{2B}^a$ and $I_\text{2B}^b$ respectively,
\be
I_\text{2B}=I_\text{2B}^a+I_\text{2B}^b.\label{eq:split2}
\ee
Figure \ref{fig:split} illustrates the splitting described in Eq. (\ref{eq:split2}).

\begin{figure}
\centering
\includegraphics[width=\linewidth]{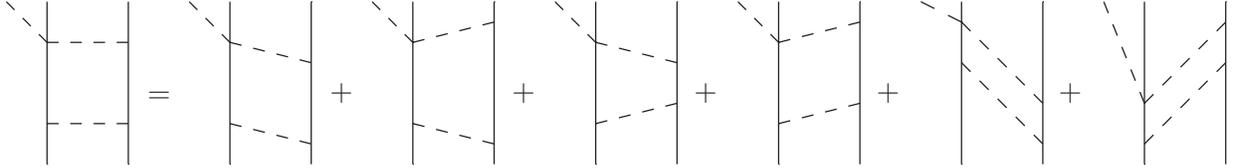}
\caption{TOPT terms resulting from the $I_\text{2B}^a$ integral.}
\label{fig:topt}
\end{figure}

Next, we perform partial fraction decomposition on each of the pion propagators, splitting each of the two terms into four terms.  Then, we continue the decomposition process until each term can be expressed as a single residue.  For $I_\text{2B}^a$ we will isolate the poles containing $\omega_f$ and then close the contour around them (for $I_\text{2B}^b$, the $\omega_i$ poles are isolated).  By isolating the poles in this way, the resulting expression is easily recognized as the sum of six TOPT terms.  For clarity, we show these terms pictorially for $I_\text{2B}^a$ in Fig. \ref{fig:topt} where we have left the overall $\frac{1}{-m_\pi}$ implicit.
We assume for now that the stretched box diagrams are small, as they were in the rescattering toy model investigation \cite{Hanhart:2000wf} and denote the sum of the four remaining terms with a $\mathring{I}$.

Finally, motivated by the interpretation which is presented in the next section, we algebraically re-combine these four terms to find
\bea
\mathring{I}_\text{2B}^a&=&\frac{1}{(m_\pi/2)^2-(\omega_f+\delta_a(\bfl))^2}\left[1+\frac{\delta_a(\bfl)}{\omega_f}\right]\frac{1}{-m_\pi}\left[1-\frac{\delta_a(\bfl)}{\omega_i+\delta_a(\bfl)}\right]\frac{1}{m_\pi-2E-\gamma}\frac{1}{-\omega_i^2}\label{eq:isifinal}
\\
\mathring{I}_\text{2B}^b&=&\frac{1}{-\omega_f^2}\frac{1}{-2E-\gamma}\left[1-\frac{\delta_b(\bfl)}{\omega_f+\delta_b(\bfl)}\right]\frac{1}{m_\pi}\left[1+\frac{\delta_b(\bfl)}{\omega_i}\right]\frac{1}{(m_\pi/2)^2-(\omega_i+\delta_b(\bfl))^2},\label{eq:fsifinal}
\eea
where we have separated out terms involving $\delta_a$ and $\delta_b$,
\bea
\delta_a(\bfl)&=&\frac{\bfl^2}{2m_N}-\frac{m_\pi}{2}+\frac{\gamma}{2}
\\
\delta_b(\bfl)&=&\frac{\bfl^2}{2m_N}+\frac{\gamma}{2},
\eea
because (as will be shown in the next section) they are sub-leading and we will neglect them in the main body of this work.  The only approximation made in the evaluation of the loop integral to obtain Eqs. (\ref{eq:isifinal}) and (\ref{eq:fsifinal}) is to neglect the stretched boxes.  Let us pause to summarize what we have done so far: (1) the DW amplitude was written down as a loop integral, (2) partial fractions was used to split the product of the two nucleon propagators into a sum $I_\text{2B}^a+I_\text{2B}^b$, (3) the loop integrals were evaluated and the result expressed in terms of TOPT diagrams, and (4) the TOPT diagrams were algebraically combined into a form useful for the following interpretation.

%%%%%%%%%%%%%%%%%%%%%%%%%%%%%%%%%%%%%%%%%%
\subsection{\label{sec:interp}Interpretation}

\begin{figure}
\begin{center}
\subfigure[\ ${\cal O}_\text{2B}^a$ of Eq. (\ref{eq:prodisi2})]{\label{fig:iaf}\includegraphics[height=1.5in]{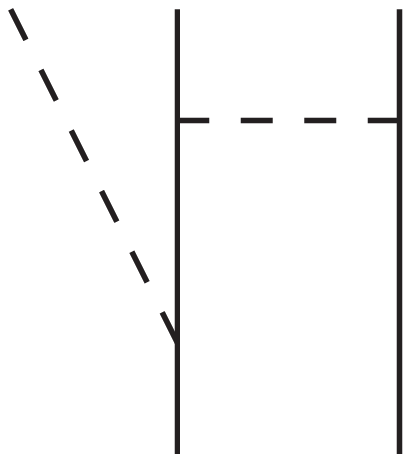}}
\hspace{.15\linewidth}
\subfigure[\ ${\cal O}_\text{2B}^b$ of Eq. (\ref{eq:prodfsi2})]{\label{fig:iai}\includegraphics[height=1.5in]{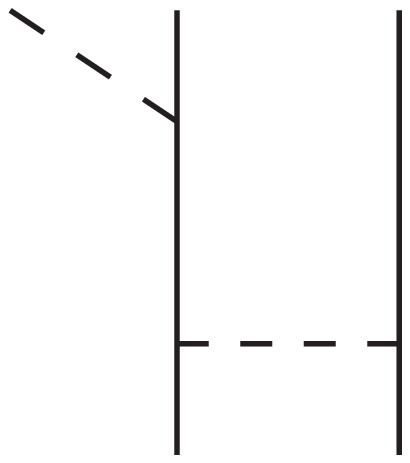}}
\end{center}
\caption{\label{fig:prodop}Two-body impulse production operators (pion exchange is non-static).}
\end{figure}

Although not obvious at first sight, convolution of the operator corresponding to Eq. (\ref{eq:isifinal}) with wave functions as defined in Eqs. (\ref{eq:mdw}) and (\ref{eq:misi}) results in an amplitude approximately equivalent to that which one obtains by using the operator shown in Fig. \ref{fig:iaf}.  The same is true of Eq. (\ref{eq:fsifinal}) with Fig. \ref{fig:iai}, and together they replace the traditional (one-body) impulse approximation with Eq. (\ref{eq:mdwfinal}).  Furthermore, the operator that results from Eq. (\ref{eq:fsifinal}) is expected to be small by power counting arguments.  The task of this subsection is to verify these statements in detail.

In Eq. (\ref{eq:isifinal}) the factor $(m_\pi-2E-\gamma)^{-1}(-\omega_i^2)^{-1}$ is interpreted as the product of the two-nucleon initial-state wave function with static OPE.  This is the statement that
\be
\frac{1}{m_\pi-2E-\gamma}\frac{1}{-\omega_i^2}=gU^\text{OPE}.
\ee
This factor can be absorbed (after adding in the PW term) using the zero-relative-energy Lippmann-Schwinger equation that is employed by the phenomenological potentials we are using.  We will continue to refer to the initial wave function as a function of $\bfp$ and $\bfp_i$, so absorbing this factor means that we set $\bfl=\bfp$.

Likewise, in Eq. (\ref{eq:fsifinal}) the factor $(-\omega_f^2)^{-1}(-2E-\gamma)^{-1}$ is interpreted as the product of static OPE with the two-nucleon final-state wave function: $U^\text{OPE}g$.  Absorbing this factor into the wave function, we set $\bfl=\bfk$.  The remaining factors of $\mathring{I}_\text{2B}^a$ and $\mathring{I}_\text{2B}^b$ become the two-body impulse production operators,
\bea
{\cal O}_\text{2B}^a&=&\frac{\bsig_1\cdot(\bfp-\bfk)\bsig_2\cdot(\bfk-\bfp)}{(m_\pi/2)^2-(\omega_f+\delta_a(\bfp))^2}\left[1+\frac{\delta_a(\bfp)}{\omega_f}\right]\frac{1}{-m_\pi}\left[1-\frac{\delta_a(\bfp)}{\omega_i+\delta_a(\bfp)}\right]\bfS\cdot\bfp\label{eq:prodisi}
\\[.1in]
{\cal O}_\text{2B}^b&=&\bfS\cdot\bfk\left[1-\frac{\delta_b(\bfk)}{\omega_f+\delta_b(\bfk)}\right]\frac{1}{m_\pi}\left[1+\frac{\delta_b(\bfk)}{\omega_i}\right]\frac{\bsig_1\cdot(\bfp-\bfk)\bsig_2\cdot(\bfk-\bfp)}{(m_\pi/2)^2-(\omega_i+\delta_b(\bfk))^2},\label{eq:prodfsi}
\eea
where we have now made explicit the momentum dependences of the vertices and used ${\bf S}=(\bsig_1+\bsig_2)/2$.  It is also important to include form factors in the OPE which match those of the wave functions.  These form factors are present in our calculation even though we leave them out of this expression for the sake of generality.

Next, note that in the evaluation of the matrix element using Eq. (\ref{eq:prodisi}), the initial state wave function is peaked about its plane wave value $\bfp\approx\bfp_i$, and thus $E\approx m_\pi/2-\gamma/2$ and $\delta_a(\bfp)\approx0$.  On the other hand, in Eq. (\ref{eq:prodfsi}), we have $\bfk\approx\bfk_i$ and $E\approx-\gamma/2$ and $\delta_b(\bfk)\approx0$.  If we were to neglect all the $\delta$'s, we would have
\bea
{\cal O}_\text{2B}^a&\approx&\frac{\bsig_1\cdot(\bfp-\bfk)\bsig_2\cdot(\bfk-\bfp)}{(m_\pi/2)^2-\left((\bfp-\bfk)^2+m_\pi^2\right)}\frac{1}{-m_\pi}\bfS\cdot\bfp\label{eq:prodisi2}
\\[.1in]
{\cal O}_\text{2B}^b&\approx&\bfS\cdot\bfk\frac{1}{m_\pi}\frac{\bsig_1\cdot(\bfp-\bfk)\bsig_2\cdot(\bfk-\bfp)}{(m_\pi/2)^2-\left((\bfp-\bfk)^2+m_\pi^2\right)},\label{eq:prodfsi2}
\eea
which suggests that these operators can be approximately interpreted as the diagrams in Fig. \ref{fig:prodop}.
Thus we have finally obtained our central result [Eq. (\ref{eq:mdwfinal})] which states that the correct impulse approximation is a two-body operator.  The contribution to pion production given in Eq. (\ref{eq:mdwfinal}) is not replacing the rescattering diagram (which is also two-body), but rather replacing the traditional contribution which has been referred to as the impulse approximation (or direct production).  Note that if we assign standard pion production power counting to these diagrams, Fig. \ref{fig:iaf} is $\mathcal{O}\left(\sqrt{\frac{m_\pi}{m_N}}\right)$ while Fig. \ref{fig:iai} is $\mathcal{O}\left(\frac{m_\pi}{m_N}\right)$.  In the next section the approximate expressions given in Eqs. (\ref{eq:prodisi2}) and (\ref{eq:prodfsi2}) are numerically evaluated.  Nevertheless, we acknowledge the importance of verifying that the $\delta$ terms are indeed small and relegate that discussion to \cref{sec:isi,sec:fsi}.

%%%%%%%%%%%%%%%%%%%%%%%%%%%%%%%%%%%%%%%
\subsection{\label{sec:evaluation}Evaluation of two-body operators}

Next, we calculate the threshold $s$-wave $np\to d\pi^0$ amplitudes corresponding to Eqs. (\ref{eq:prodisi2}) and (\ref{eq:prodfsi2}).  We do not present the details here as most are given in Ref. \cite{Bolton:2010qu}.  Again, we remind the reader that for the sake of consistency we use a deuteron wave function calculated from a purely-OPE potential (with form factors as described in Appendix \ref{sec:deuteron}).  For the initial-state distorted waves, we use three different phenomenological potentials (Av18 \cite{Wiringa:1994wb}, Nijmegen II \cite{Stoks:1994wp}, and Reid `93 \cite{Stoks:1994wp}).  In Table \ref{tab:results}, we display the results in terms of the reduced matrix elements of Ref. \cite{Bolton:2009rq}.
\begin{table}
\caption{\label{tab:results}Threshold reduced matrix elements calculated with an OPE deuteron and various phenomenological initial states.  The first row shows the traditional impulse approximation (one-body) while the second and third show our replacement (two-body).}
\begin{center}
\renewcommand{\tabcolsep}{3mm}
\begin{tabular}{cccc}
\hline\hline
& Av18 & Reid '93 & Nijm II\\ \hline
$A^\text{DW}_\text{1B}$ & 8.3 & 7.1 & 5.4\\
$A_\text{2B}^\text{DW,a}$ & 17.4 & 13.5 & 7.8\\
$A_\text{2B}^\text{DW,b}$ & $-$1.5 & $-$2.2 & $-$6.9\\ \hline\hline
\end{tabular}
\end{center}
\end{table}

The first row of Table \ref{tab:results} gives the traditional (one-body) impulse approximation, which is slightly bigger than Ref. \cite{Bolton:2010qu} due to the use of the OPE deuteron.  The next row shows that the new two-body operator (at leading order) is roughly twice as large as the traditional calculation it is replacing.  We mention here that the significant cancellation between deuteron $s$- and $d$-states remains, keeping the impulse amplitude smaller than rescattering; however, the cancellation is less complete when using our new two-body operator.  The final row verifies that the ${\cal O}_\text{2B}^b$ diagram is smaller than the ${\cal O}_\text{2B}^a$ diagram, as dictated by the power counting.  The Nijmegen II potential provides a bit of deviation from these results, and it will be interesting to investigate other potentials to determine the true model dependence of this calculation.  In finding these results, it is important that the pion propagators of Eqs. (\ref{eq:prodisi2}) and (\ref{eq:prodfsi2}) be implemented in a manner consistent with the potential used for the wave function of Fig. \ref{fig:loop}.  Namely, the cutoff procedure of the convolution integral with form factors needs to match that by which the potential was constructed.  Appendix \ref{sec:deuteron} contains the details of this procedure.

Our conclusion is that the traditional impulse approximation is an underestimate.  While it is true that several approximations were made in order to permit final expressions as simple as Eqs. (\ref{eq:prodisi2}) and (\ref{eq:prodfsi2}), we believe this conclusion to be sound.  The $\delta$ terms do not defy their classification as sub-leading (see \cref{sec:isi,sec:fsi}), and \cref{sec:sigma} shows that using $K=\text{OPE}$ in the initial state is at least reasonable.  In summary, we simply claim that Eq. (\ref{eq:prodisi2}) is the new impulse approximation at leading order in the effective field theory.  The corrections in the aforementioned appendices, in addition to Eq. (\ref{eq:prodfsi2}) contribute to the next-to-leading order calculation, which needs to be systematically considered in a later work.

Finally, it is important to note that although the OPE deuteron reproduces the phenomenological results for the rescattering diagram quite well, the numbers in this section are greatly changed if a phenomenological deuteron is used.  Using Av18 we find  $A^\text{DW}_\text{1B}=4.9$, and by using the cutoff procedure of Av18 for the two-body operators, we find $A_\text{2B}^\text{DW,a}=33.5$, $A_\text{2B}^\text{DW,b}=-2.8$.  Thus, the ratio of our new two-body operator to the traditional impulse operator is $\sim7$ instead of the $\sim2$ presented above.  At this time one is faced with a choice of either: (1) using a \lqt correct" phenomenological deuteron and leaving out parts of the potential when calculating the two-body kernel or (2) using an inexact OPE deuteron with a completely self-consistent kernel.  For the time being, we believe the latter to be more trustworthy, if not ideal.

%%%%%%%%%%%%%%%%%% DISCUSSION %%%%%%%%%%%%%%%%%%%
\section{\label{sec:discussion}Discussion}

Experimental data for pion production near threshold are reported in terms of two parameters, $\alpha$ and $\beta$, defined for $np\rightarrow d\pi^0$,
\be
\sigma(\eta)=\frac{1}{2}\left(\alpha\eta+\beta\eta^3\right),
\ee
where $\eta$ is the pion momentum in units of its mass.  Table VI of Ref. \cite{Bolton:2010qu} shows the results obtained by the four most recent experiments.  Since the present calculation is performed at threshold ($\eta=0$), we compute only the value of $\alpha$,
\be
\alpha=\frac{m_\pi}{128\pi^2sp}\left|A\right|^2,\label{eq:alpha}
\ee
where $s=(m_d+m_\pi)^2$ is the square of the invariant energy.  For ease of comparison, we invert Eq. (\ref{eq:alpha}), plug in the results of the mentioned experiments, and propagate the errors to find Table \ref{tab:expt}.
\begin{table}
\caption{\label{tab:expt}Threshold reduced matrix elements extracted from experiment}
\begin{center}
\renewcommand{\tabcolsep}{3mm}
\begin{tabular}{cc}
\hline\hline
Experiment & $A^\text{expt}$\\ \hline
$np\rightarrow d\pi^0$ \cite{Hutcheon:1989bt} & $80.1\pm1.1$\\
$\vec{p}p\rightarrow d\pi^+$ (Coulomb corrected) \cite{Heimberg:1996be} & $85.2\pm1.0$\\
$pp\rightarrow d\pi^+$ (Coulomb corrected) \cite{Drochner:1998ja} & $84.6\pm1.9$\\
Pionic deuterium decay \cite{Strauch:2010rm} & $93.8^{+0.9}_{-2.0}$\\ \hline\hline
\end{tabular}
\end{center}
\end{table}

The full theoretical amplitude includes not only the impulse diagram but also the rescattering diagram, which is given in Table \ref{tab:tot} along with the total amplitude using either the traditional one-body or the leading-order two-body impulse diagram.
\begin{table}
\caption{\label{tab:tot}Rescattering (RS) and total reduced matrix elements for a variety of potentials.  The second line shows the traditional calculation (with a one-body IA) while the third shows our replacement (with a two-body IA).}
\begin{center}
\renewcommand{\tabcolsep}{3mm}
\begin{tabular}{cccc}
\hline\hline
& Av18 & Reid '93 & Nijm II\\ \hline
RS & 69.8 & 72.1 & 74.0\\
RS + IA (1B) & 78.1 & 79.2 & 79.4\\
RS + IA (2B) & 87.2 & 85.6 & 81.8\\ \hline\hline
\end{tabular}
\end{center}
\end{table}
The uncertainty in an effective field theory calculation is estimated by the power counting scheme.  In this work, we have included both the rescattering and the impulse diagrams up to ${\cal O}\left(\sqrt{m_\pi/m_N}\right)$.  Therefore one might assign an uncertainty of $m_\pi/m_N=14\%$ to the calculation but stress that such an estimate based solely on power counting is rough at best.  Taking this uncertainty, we see that the theory update presented here changes the situation from under-prediction of the most recent pionic deuterium experiment by $\sim1.3\,\sigma$, to under-prediction by $\sim0.7\,\sigma$.

In summary, we have developed a consistent formalism that allows one to separate effects of the kernel from those of the wave functions, finding a new impulse approximation kernel.  This two-body operator, given in Eq. (\ref{eq:mdwfinal}), replaces the traditional one-body impulse approximation and is the central result of the present work.  We numerically investigated the simplest example ($s$-wave $NN\rightarrow d\pi$) and found the impulse amplitude to be increased by a factor of roughly two over the tradational amplitude.  This calculation was performed with a regulated OPE deuteron which has advantages and disadvantages as described in the body of this work.  Rescattering remains the dominant contribution to the cross section.  We find that the updated total cross section is $\sim10\%$ larger than before and is in agreement with experiment at leading order.  We verified that corrections to the new impulse approximation (which together with other loops and counterterms will contribute at next-to-leading order) do not destroy these results.

These findings suggest several directions for future research.  Firstly, one needs to develop a power counting scheme for the \lqt Q space" discussed in Sec. \ref{sec:bs}.  Secondly, the significant model dependence of the new formulation of the impulse approximation needs to be investigated in a renormalization group invariant way.  Thirdly, it will be very interesting to see the impact of this increased impulse amplitude on the $pp\rightarrow pp\pi^0$ cross section which is suppressed due to the absence of rescattering.  Finally, one could look at the energy dependence ($p$-wave pions) of $NN\rightarrow NN\pi$, for which there is an abundance of experimental data.

\begin{acknowledgments}
This research was supported in part by the U.S. Department of Energy.  We thank Christoph Hanhart and Vadim Baru for multiple valuable conversations and for suggesting the investigation of the full loop integral.
\end{acknowledgments}

\begin{appendix}

%%%%%%%%%%%%%%%%% LAGRANGIAN %%%%%%%%%%%%%%%%%%%%%%
\section{Lagrange densities\label{sec:lagrangian}}

We define the index of a Lagrange density to be
\be
\nu=d+\frac{f}{2}-2\label{eq:index},
\ee
where $d$ is the sum of the number of derivatives and powers of $m_\pi$, and $f$ is the number of fermion fields.  This represents the standard power counting for nuclear physics.  The $\nu=0$ Lagrangian (with spatial vectors in bold font) is \cite{Cohen:1995cc}
\be
\mathcal{L}^{\left(0\right)}=\frac{1}{2}\left(\partial\pi_a\right)^2-\frac{1}{2}m_\pi^2\pi_a^2+N^\dagger i\partial_0N+\frac{g_A}{2f_\pi}N^\dagger\left(\tau_a\bsig\cdot\bnabla\pi_a\right)N+...,\label{eq:l0}
\ee
where $\tau_a$ and $\bsig$ are the Pauli matrices acting on the isospin and spin of a single nucleon.  The \lqt$+...$" indicates that only the terms used in this calculation are shown.

The $\nu=1$ Lagrangian includes recoil corrections and other terms invariant under SU(2)$_L\times\,$SU(2)$_R$.
\be
\mathcal{L}^{\left(1\right)}=\frac{1}{2m_N}N^\dagger\bnabla^2N-\frac{1}{2m_N}\left[\frac{g_A}{2f_\pi}iN^\dagger\tau_a\dot{\pi}_a\bsig\cdot\bnabla N+H.c.\right]+...,\label{eq:l1}
\ee
where we use the values given in Table \ref{tab:parameters}.
\begin{table}
\caption{Parameters used.\label{tab:parameters}}
\begin{center}
\renewcommand{\tabcolsep}{3mm}
\begin{tabular}{cc}
\hline\hline
$m_\pi=134.98\ \text{MeV}$ & $g_A=1.32\ \text{MeV}$\\
$m_N=938.92\ \text{MeV}$ & $f_\pi=92.4\ \text{MeV}$\\ \hline\hline
\end{tabular}
\end{center}
\end{table}
Note that the terms with the $c_i$ low energy constants which appear at this order do not get promoted in MCS for these kinematics and are thus not used.  Also, the terms with the $d_i$ low energy constants do not contribute to s-wave pion production.  Finally, the $NNNN$ contact terms, $C_{S,T}$, do not contribute because we are using a potential with a repulsive core [$R_i(r)R_f(r)\rightarrow0$ as $r\rightarrow0$ for $l_i=1$, $l_f=0$].

%%%%%%%%%%%%%%%%%%%%%% NNpi %%%%%%%%%%%%%%%%%%%%%
\section{$N\rightarrow N\pi$ from BChPT\label{sec:nnpi}}

The LO $NN\pi$ interaction reads,
\be
\mathcal{L}^{(0)}\subset\overline{N}\frac{g_A}{2}\slashed{u}_\perp\gamma^5N,
\ee
where $u_{\perp,\mu}=u_\mu-v\cdot u\,v_\mu$, $u_\mu=i(u^\dagger\partial_\mu u-u\partial_\mu u^\dagger)$, and $u^2=e^{i\tau_a\pi_a/f_\pi}$.  We find
\bea
u_\mu&=&i(i\tau_a\partial_\mu\frac{\pi_a}{2f_\pi}-(-i)\tau_a\partial_\mu\frac{\pi_a}{2f_\pi})=-\frac{\tau_a}{f_\pi}\partial_\mu\pi_a\nonumber
\\
\slashed{u}_\perp&=&\gamma_0(u_0-u_0\cdot1)-\gamma_i\left(-\frac{\tau_a}{f_\pi}\partial_i\pi_a\right)\qquad\qquad i=1,2,3\nonumber
\\
&=&\frac{\tau_a}{f_\pi}\gamma_i\partial_i\pi_a,
\eea
and thus the Feynman rule for an outgoing pion with momentum $q$ and isospin $a$ is
\be
{\cal O}_\pi^{(0)}=-i\left(\frac{g_A}{2f_\pi}\gamma_i\gamma^5\right)\left(iq_i\right)\tau_a.
\ee

At threshold, the pion four-momentum is $q=(m_\pi,0,0,0)$ making ${\cal O}_\pi^{(0)}=0$.  This reflects the fact (well-known from current algebra) that threshold pion production proceeds via the off-diagonal, and therefore $1/m_N$ suppressed, interaction $g_\pi\gamma^5\gamma^0q^0\tau$.  In the effective theory, this recoil correction shows up in the NLO Lagrangian
\be
\mathcal{L}^{(1)}\subset-i\frac{g_A}{2m_N}\overline{N}\left\{v^\mu u_\mu,S^\mu\partial_\mu\right\}N,
\ee
where the spin vector is $S^\mu=-\frac{1}{2}\gamma^5(\gamma^\mu\slashed{v}-v^\mu)$.  Thus the Feynman rule is
\begin{align}\begin{split}
\mathcal{O}_\pi^{(1)}&=-i\left(-i\frac{g_A}{2m_N}\right)\left[-\frac{\tau_a}{f_\pi}(-im_\pi)\right]\left[\frac{1}{2}\gamma^5\gamma_i\gamma^0(\overrightarrow{\bnabla}-\overleftarrow{\bnabla})_i\right]
\\
&=-i\frac{m_\pi}{2m_N}\frac{g_A}{2f_\pi}\gamma^5\gamma_i\gamma^0(\overrightarrow{\bnabla}-\overleftarrow{\bnabla})_i\tau_a,\label{eq:feynrule}
\end{split}\end{align}
where the derivatives act on the nucleon wave functions.

%%%%%%%%%%%%%%%%%%%%% OPE DEUTERON %%%%%%%%%%%%%%%%%%
\section{\label{sec:deuteron}One pion exchange deuteron}

\begin{figure}
\centering
\includegraphics[height=2in]{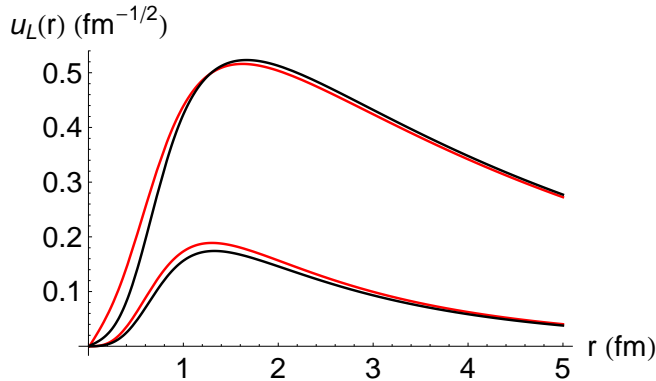}
\caption{\label{fig:opedeuteron}(Color online) Deuteron $s$- and $d$-state wave functions (the $s$-state is larger).  The potentials used to calculate the wave functions are Av18 (black) and the cutoff OPE described in this section (red).}
\end{figure}

In this Appendix, we present the method by which the deuteron wave function is calculated for use in Sec. \ref{sec:distortions}.  This method is taken directly from the work of Friar, Gibson, and Payne \cite{Friar:1984wi}.  The OPE potential is defined to have central ($Y$) and tensor ($T$) parts,
\be
V_\pi(\bfr)=f^2m_\pi\frac{\tau_{1,a}\tau_{2,a}}{3}\left[\bsig_1\cdot\bsig_2Y(r)+S_{12}T(r)\right],\label{eq:opedef}
\ee
where $f^2=0.079$ (to be distinguished from $f_\pi$) measures the strength of the pion-nucleon coupling and $S_{12}$ is the standard tensor operator.  The deuteron has isospin zero and spin one, so we have
\be
V_\pi(\bfr)=-f^2m_\pi\left[Y(r)+S_{12}T(r)\right].
\ee
The $Y$ and $T$ functions are expressed as derivatives of the Fourier transform of the pion propagator,
\bea
Y(r)&=&h_0''(x)-h_0'(x)/x\nonumber
\\
T(r)&=&h_0''(x)+2h_0'(x)/x\nonumber
\\
h_0(x)&=&\frac{4\pi}{(2\pi)^3m_\pi}\int d^3q\frac{e^{-i\bfq\cdot\bfr}}{\bfq^2+m_\pi^2}F^2_{\pi NN}(\bfq^2)\label{eq:h0}
\eea
where $x=m_\pi r$ and $F_{\pi NN}$ is the form factor for which we use,
\be
F_{\pi NN}(\bfq^2)=\left(\frac{\Lambda^2-m_\pi^2}{\bfq^2+\Lambda^2}\right)^n.
\ee
In Ref. \cite{Friar:1984wi}, it is shown that
\bea
Y(r)&=&\frac{e^{-x}}{x}-\beta^3e^{-\beta x}\sum_{i=0}^{2n-1}\frac{\xi^i}{i!}\left(\delta_i(\beta x)-2i\delta_{i-1}(\beta x)\right)
\\
T(r)&=&\frac{e^{-x}}{x}\left(1+\frac{3}{x}+\frac{3}{x^2}\right)-\beta^3e^{-\beta x}\sum_{i=0}^{2n-1}\frac{\xi^i}{i!}\left[\delta_i(\beta x)-(2i-3)\delta_{i-1}(\beta x)\right]
\eea
where $\beta=\Lambda/m_\pi$ and $\xi=(\beta^2-1)/2\beta^2$ and the $\delta_i$ are defined by
\be
\delta_{i+1}(\beta x)=(2i-1)\delta_i(\beta x)+(\beta x)^2\delta_{i-1}(\beta x)
\ee
along with $\delta_0=1/\beta x$ and $\delta_1=1$.  One of the results of Ref. \cite{Friar:1984wi} is that larger values of $n$ lead to better fits to experimental data.  We use $n=5$ and $\beta/\sqrt{10}=5.687805$ in order to precisely reproduce the binding energy $E_B=2.2246$ MeV.  The wave functions are calculated by integrating in from $r_\text{max}=100$ fm and adding together two linearly independent solutions such that the sum vanishes at $r_\text{min}=0.01$ fm.  As shown in Fig. \ref{fig:opedeuteron}, the results are close to the \lqt correct" Av18 deuteron.

In Table \ref{tab:deuteron}, we display the quadrupole moment and mean square charge radius of Av18, this OPE potential, and experiment (as quoted in \cite{Friar:1984wi}).
\begin{table}
\caption{\label{tab:deuteron}Deuteron properties.}
\begin{center}
\renewcommand{\tabcolsep}{3mm}
\begin{tabular}{ccc}
\hline\hline
Potential & Q (fm$^2$) &  $\la r^2\ra^{1/2}$ (fm)\\ \hline
Av18 & 0.270 & 1.968\\
OPE ($n=5$) & 0.282 & 1.939\\
Experiment & 0.2859(3) & 1.955(5)\\ \hline\hline
\end{tabular}
\end{center}
\end{table}
It is clear that the form factors in the OPE potential make it difficult to distinguish this construction as less accurate than Av18.  Finally, in Table \ref{tab:rs}, we display the reduced matrix elements for the rescattering pion production diagram evaluated with both the phenomenological potentials and the deuteron of this section.  Since this diagram makes the largest contribution to the cross section we need to verify that neglecting non-OPE parts of the potential does not dramatically change this amplitude.
\begin{table}
\caption{\label{tab:rs}Effect of using OPE deuteron on rescattering diagram.}
\begin{center}
\renewcommand{\tabcolsep}{3mm}
\begin{tabular}{cccc}
\hline\hline
Deuteron & Av18 & Reid '93 & Nijm II\\ \hline
Phenomenological & 67.8 & 69.7 & 71.1\\
OPE ($n=5$) & 69.8 & 72.1 & 74.0\\ \hline\hline
\end{tabular}
\end{center}
\end{table}
Indeed, we observe what should be expected: since the rescattering diagram is not as sensitive to the core of the deuteron, using the OPE wave function in place of the standard one has only a small effect.

%%%%%%%%%%%%%%%%%%%%%%%%%%%%%%%%%%%%%%%%
\section{\label{sec:isi}Effect of the $\delta$ terms: ${\cal O}_\text{2B}^a$ diagram}

In this section we calculate the correction terms to the first two-body DW amplitude [Eq. (\ref{eq:prodisi})] which is shown in Fig. \ref{fig:iaf}.  Assuming that the $\delta$'s truly are small compared to Eq. (\ref{eq:prodisi2}), we will only worry about calculating them one at a time, numbering the contribution of the $\delta$'s from right to left as 1, 2 and 3.  Note that we will display the results as calculated using the OPE deuteron and the Av18 initial state.

\subsection{\label{sec:isi1}First ${\cal O}_\text{2B}^a$ correction term: $\Delta{\cal O}_1$}
\begin{figure}
\begin{center}
\includegraphics[height=1.5in]{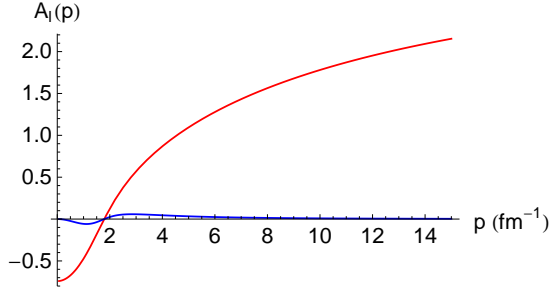}
\caption{\label{fig:Al}(Color online) Coefficients of the expansion in Eq. (\ref{eq:sphexp}).  The red curve shows $l=0$ and the blue shows $l=2$.}
\end{center}
\end{figure}

Consider the rightmost $\delta$ in Eq. (\ref{eq:prodisi}),
\bea
\Delta{\cal O}_1&=&-\frac{\bsig_1\cdot(\bfp-\bfk)\bsig_2\cdot(\bfk-\bfp)}{-(\bfp-\bfk)^2-\mu^2}F^2_{\pi NN}((\bfp-\bfk)^2)\frac{1}{-m_\pi}\nonumber
\\
&&\qquad\qquad\times\frac{\frac{\bfp^2}{2m_N}-\frac{m_\pi}{2}+\frac{\gamma}{2}}{\sqrt{(\bfp-\bfp_i)^2+m_\pi^2}+\frac{\bfp^2}{2m_N}-\frac{m_\pi}{2}+\frac{\gamma}{2}}\bfS\cdot\bfp,\label{eq:blah}
\eea
where $\mu^2=3m_\pi^2/4$ and $F_{\pi NN}$ is the form factor described in \cref{sec:deuteron}.  The easiest way to evaluate the matrix element of this operator is to let the OPE act to the left on the deuteron in position space.  The resulting expression is then transformed to momentum space.  We can expand the fraction in the integrand of Eq. (\ref{eq:blah}) into spherical harmonics (taking $\hat{\bfp}_i=\hat{{\bf z}}$),
\be
\frac{\frac{\bfp^2}{2m_N}-\frac{m_\pi}{2}+\frac{\gamma}{2}}{\sqrt{(\bfp-\bfp_i)^2+m_\pi^2}+\frac{\bfp^2}{2m_N}-\frac{m_\pi}{2}+\frac{\gamma}{2}}=\sum_{l}A_{l}(p)Y_{l,0}(\hat{\bfp}),\label{eq:sphexp}
\ee
and note that only the $l=0,2$ terms will contribute to $s$-wave production.  The expansion coefficients are shown in Fig. \ref{fig:Al}.
Clearly the $l=2$ term is small and we neglect it here to avoid the extra algebra involved with a $J=2$ operator (resulting in the $A_2$ reduced matrix elements in the notation of Ref. \cite{Bolton:2009rq}).  We find
\be
\frac{\Delta{\cal M}_1}{{\cal M}}=-34\%.
\ee

\subsection{\label{sec:isi2}Second ${\cal O}_\text{2B}^a$ correction term: $\Delta{\cal O}_2$}
\begin{figure}
\centering
\parbox{.4\linewidth}{\includegraphics[width=\linewidth]{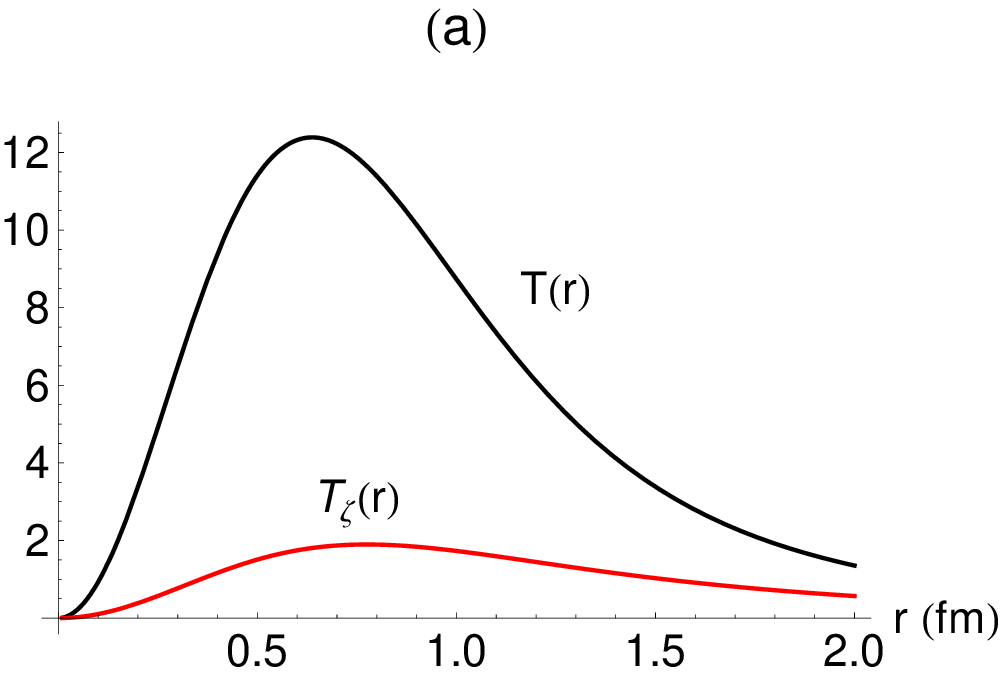}}
\hspace{.05\linewidth}
\parbox{.4\linewidth}{\includegraphics[width=\linewidth]{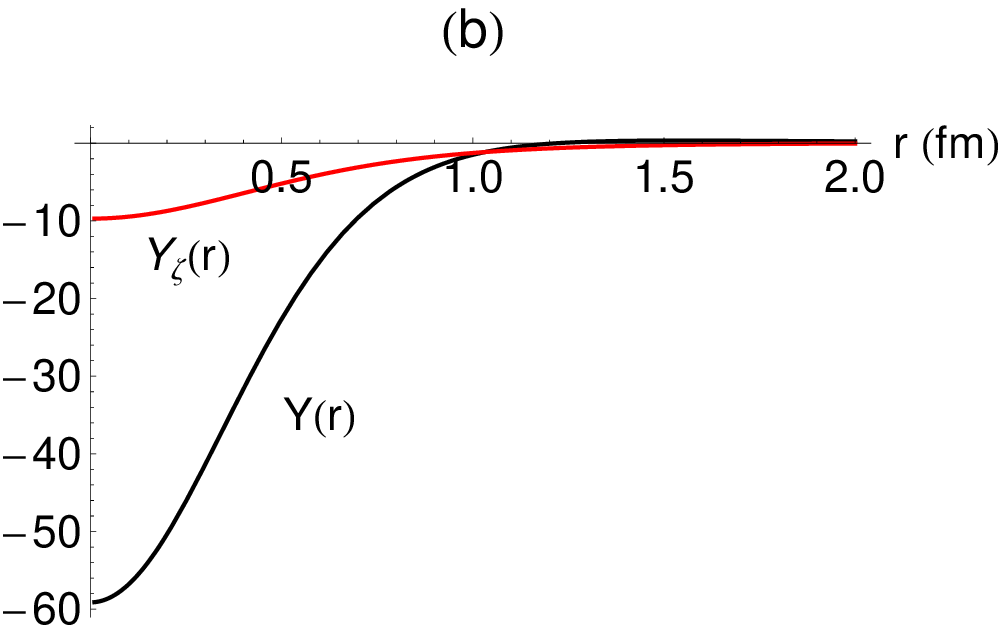}}
\caption{\label{fig:zeta}(Color online) Effect of the square root on the OPE a) tensor and b) central radial functions.}
\end{figure}

Next consider the term,
\bea
\Delta{\cal O}_2&=&\frac{\bsig_1\cdot(\bfp-\bfk)\bsig_2\cdot(\bfk-\bfp)}{-(\bfp-\bfk)^2-\mu^2}F_{\pi NN}^2((\bfp-\bfk)^2)\frac{1}{\sqrt{(\bfp-\bfk)^2+m_\pi^2}}\frac{1}{-m_\pi}\nonumber
\\
&&\qquad\qquad\times\left(\frac{\bfp^2}{2m_N}-\frac{m_\pi}{2}+\frac{\gamma}{2}\right)\bfS\cdot\bfp.
\eea
This term has a modified OPE,
\bea
\lefteqn{\int\frac{d^3q}{(2\pi)^3}e^{-i\bfq\cdot\bfr}\frac{\bsig_1\cdot\bfq\,\bsig_2\cdot(-\bfq)}{\bfq^2+\mu^2}F_{\pi NN}^2(\bfq^2)\frac{1}{\sqrt{\bfq^2+m_\pi^2}}}\label{eq:sqrt}
\\
&\equiv&\bsig_1\cdot\nabla\bsig_2\cdot\nabla\frac{\zeta(r)}{4\pi}
\\
&=&\frac{\mu^2}{12\pi}\left[S_{12}T_\zeta(r)+\bsig_1\cdot\bsig_2Y_\zeta(r)\right].
\eea
In Fig. \ref{fig:zeta} we compare the functions $T_\zeta(r)$ and $Y_\zeta(r)$ to traditional OPE which has $\mu$ in place of the square root in Eq. (\ref{eq:sqrt}).
We use the Schr\"odinger equation to replace the $\bfp^2/2m_N$ with $V(r)$ and evaluate the matrix element in position space to find
\be
\frac{\Delta{\cal M}_2}{{\cal M}}=+50\%.
\ee

\subsection{\label{sec:isi3}Third ${\cal O}_\text{2B}^a$ correction term: $\Delta{\cal O}_3$}
\begin{figure}
\centering
\parbox{.4\linewidth}{\includegraphics[width=\linewidth]{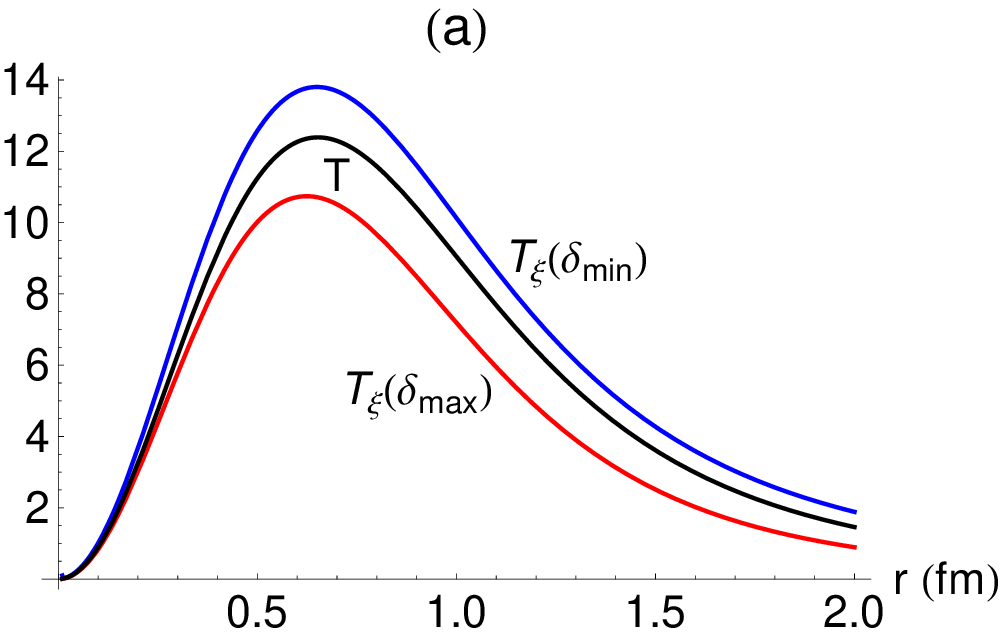}}
\hspace{.05\linewidth}
\parbox{.4\linewidth}{\includegraphics[width=\linewidth]{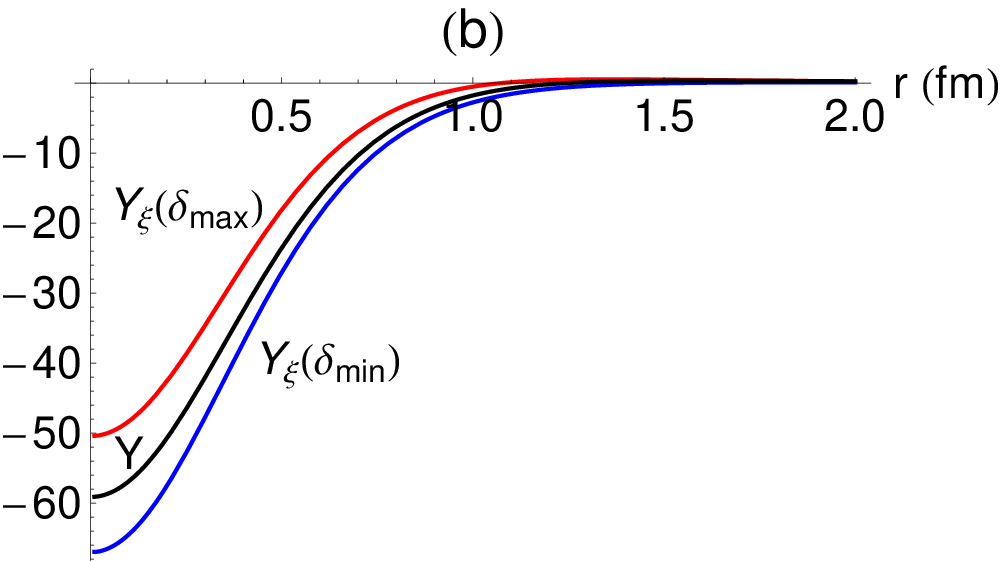}}
\caption{\label{fig:xiplots}(Color online) Effect of the $\delta$ on the OPE (a) tensor and (b) central radial functions.}
\end{figure}

Calculating the effects of the $\delta$ in the denominator of the OPE is difficult to do exactly due to the combination of momenta that appear,
\be
\Delta{\cal O}_3=\frac{\bsig_1\cdot(\bfp-\bfk)\bsig_2\cdot(\bfk-\bfp)}{-\left(\sqrt{(\bfp-\bfk)^2+m_\pi^2}+\delta(\bfp)\right)^2+m_\pi^2/4}F^2_{\pi NN}((\bfp-\bfk)^2)\frac{1}{-m_\pi}\bfS\cdot\bfp,
\ee
(recall that $\delta(\bfp)=\bfp^2/2m_N-m_\pi/2+\gamma/2$).  Instead we will evaluate it for fixed values of $\delta$ which represent the deviation of $\bfp$ away from $\bfp_i$,
\bea
\delta_+&=&\delta(p_i+m_\pi)=\frac{2p_im_\pi+m_\pi^2}{2m_N}=0.45\,m_\pi
\\
\delta_-&=&\delta(p_i-m_\pi)=\frac{-2p_im_\pi+m_\pi^2}{2m_N}=-0.32\,m_\pi.
\eea
The modified tensor and central functions $T_\xi$ and $Y_\xi$ are shown in Fig \ref{fig:xiplots}.
We define the correction as
\be
\Delta{\cal M}_3(\delta)={\cal M}(\delta)-{\cal M}(0).
\ee
We find,
\bea
\frac{\Delta{\cal M}_3(\delta_+)}{{\cal M}}&=&+16\%
\\[.1in]
\frac{\Delta{\cal M}_3(\delta_-)}{{\cal M}}&=&-32\%.
\eea

\subsection{\label{sec:isisummary}Summary of ${\cal O}_\text{2B}^a$ corrections}
For the purposes of estimating the net result we take the average of the estimates in Sec. \ref{sec:isi3} and find
\be
\frac{\Delta{\cal M}_\text{tot}}{{\cal M}}\approx-34\%+50\%-8\%=+8\%.
\ee
We have successfully shown that the corrections to the first two-body DW amplitude are small, and actually \textit{increase} the amplitude which is already twice as large as the traditional impulse approximation.

%%%%%%%%%%%%%%%%%%%%%%%%%%%%%%%%%%%%%%%
\section{\label{sec:fsi}Effect of the $\delta$ terms: ${\cal O}_\text{2B}^b$ diagram}

The second two-body DW amplitude's corrections are evaluated exactly as in the previous sub-sections and we just display the results here,
\bea
\frac{\Delta{\cal M}_1}{{\cal M}}&=&-35\%
\\[.1in]
\frac{\Delta{\cal M}_2}{{\cal M}}&=&-39\%
\\[.1in]
\frac{\Delta{\cal M}_3(\delta_+)}{{\cal M}}&=&\frac{\Delta{\cal M}_3(\delta_-)}{{\cal M}}=+3\%,
\eea
with the net result
\be
\frac{\Delta{\cal M}_\text{tot}}{{\cal M}}\approx-35\%-39\%+3\%=-71\%.
\ee
We see that the corrections to the approximation in Eq. (\ref{eq:prodfsi2}) are fairly large, but this has a negligible effect because the amplitude is already small compared to the first two-body DW amplitude.

%%%%%%%%%%%%%%%%%%%%%%%%%%%%%%%%%%%%%%%%%%
\section{\label{sec:sigma}Heavy meson exchange}

\begin{figure}
\begin{center}
\includegraphics[height=1.5in]{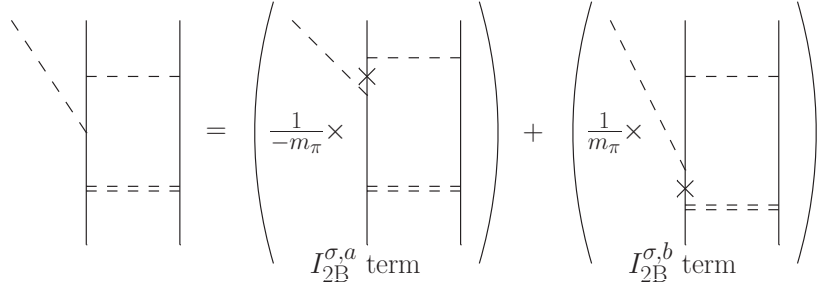}
\caption{\label{fig:sig}Impulse approximation with distorted waves: initial state heavy meson exchange.  Solid lines represent nucleons, dashed lines pions, and the double solid line a $\sigma$ meson.  Crosses represent propagators which are absent due to the partial fractions decomposition.}
\end{center}
\end{figure}

Consider the loop on the left-hand side of Fig. \ref{fig:sig} which is obtained by using OPE for the left $K$ in Eq. (\ref{eq:misi}) and $\sigma$ exchange (the dominant intermediate-range mechanism) for the right $K$.
Note that this loop only differs from Fig. \ref{fig:loop} in two ways: the meson-nucleon vertex (here we consider only scalar-isoscalar) and the meson mass.  We use a typical set of parameters \cite{Ericson:1988gk}, $g^2_\sigma/4\pi=7.1$ and $m_\sigma=550$ MeV.

The result of integrating over energy will proceed exactly as it did with the pion resulting in the two diagrams shown on the right-hand side of Fig. \ref{fig:sig}.
To interpret the $I_\text{2B}^{\sigma,a}$ term (again, neglecting stretched box diagrams), we absorb the sigma exchange into the initial state and no new term is added.  However, in the $I_\text{2B}^{\sigma,b}$ term, after absorbing the pion exchange into the final state, we are left with a new operator.  The amplitude for this operator can be obtained from that of Fig. \ref{fig:iai} with the following change:
\be
\left(\frac{g_A}{2f_\pi}\right)^2\frac{\mu^3}{3}\left[2\left(1+\frac{3}{\mu r}+\frac{3}{(\mu r)^2}\right)+1\right]\frac{e^{-\mu r}}{\mu r}\rightarrow g_\sigma^2\mu_\sigma\frac{e^{-\mu_\sigma r}}{\mu_\sigma r},\label{eq:change}
\ee
where $\mu^2=3m_\pi^2/4$ and $\mu_\sigma^2=m_\sigma^2-(m_\pi/2)^2$.  We find,
\be
A_\text{2B}^{\sigma,b}=-7.24,
\ee
which is larger in magnitude than the pionic $A_\text{2B}^{\text{DW},b}$ (with the same sign) but smaller than $A_\text{2B}^{\text{DW},a}$ (with the opposite sign).  Since $m_\sigma$ is relatively large, we can safely ignore the two $\delta$ corrections that are competing with $\omega_i$ and only need to evaluate
\be
\frac{\Delta{\cal M}_1}{{\cal M}}=-27\%.
\ee

One natural question is whether the static $\sigma$ exchange already present in the initial-state wave function is a sufficient approximation for the contribution considered in this section.  To answer this question, we can evaluate the traditional impulse approximation with
\be
|\Psi\ra_i^\sigma=|p_1,p_2\ra+GV_\sigma|\Psi\ra_i,
\ee
where here we employ a static $\sigma$ exchange that is present (at least effectively) in the wave function.  Using this initial-state wave function, we calculate
\be
{\cal M}_\text{1B}^\sigma={}_f\la\phi|{\cal O}_\pi |\Psi\ra_i^\sigma,
\ee
and find the reduced matrix element,
\be
A_\text{1B}^\sigma=-3.3.
\ee
Thus we see that the $\sigma$ exchange in the traditional impulse approximation is an underestimate (in magnitude) of the true non-static exchange dictated by the loop integral.

Of course there is no $\sigma$ in traditional B$\chi$PT, so this section is simply telling us that to achieve high accuracy it is indeed important to use more than just simple pion exchange when forming the original box diagram.  Such a calculation is beyond the scope of this work.

\end{appendix}

\bibliography{referencesPRC}

\begin{thebibliography}{33}
\expandafter\ifx\csname natexlab\endcsname\relax\def\natexlab#1{#1}\fi
\expandafter\ifx\csname bibnamefont\endcsname\relax
  \def\bibnamefont#1{#1}\fi
\expandafter\ifx\csname bibfnamefont\endcsname\relax
  \def\bibfnamefont#1{#1}\fi
\expandafter\ifx\csname citenamefont\endcsname\relax
  \def\citenamefont#1{#1}\fi
\expandafter\ifx\csname url\endcsname\relax
  \def\url#1{\texttt{#1}}\fi
\expandafter\ifx\csname urlprefix\endcsname\relax\def\urlprefix{URL }\fi
\providecommand{\bibinfo}[2]{#2}
\providecommand{\eprint}[2][]{\url{#2}}

\bibitem[{\citenamefont{Weinberg}(1979)}]{Weinberg:1978kz}
\bibinfo{author}{\bibfnamefont{S.}~\bibnamefont{Weinberg}},
  \bibinfo{journal}{Physica} \textbf{\bibinfo{volume}{A96}},
  \bibinfo{pages}{327} (\bibinfo{year}{1979}).

\bibitem[{\citenamefont{Gasser and Leutwyler}(1984)}]{Gasser:1983yg}
\bibinfo{author}{\bibfnamefont{J.}~\bibnamefont{Gasser}} \bibnamefont{and}
  \bibinfo{author}{\bibfnamefont{H.}~\bibnamefont{Leutwyler}},
  \bibinfo{journal}{Ann. Phys.} \textbf{\bibinfo{volume}{158}},
  \bibinfo{pages}{142} (\bibinfo{year}{1984}).

\bibitem[{\citenamefont{Bernard et~al.}(1993)\citenamefont{Bernard, Kaiser, and
  Meissner}}]{Bernard:1993nj}
\bibinfo{author}{\bibfnamefont{V.}~\bibnamefont{Bernard}},
  \bibinfo{author}{\bibfnamefont{N.}~\bibnamefont{Kaiser}}, \bibnamefont{and}
  \bibinfo{author}{\bibfnamefont{U.~G.} \bibnamefont{Meissner}},
  \bibinfo{journal}{Z. Phys.} \textbf{\bibinfo{volume}{C60}},
  \bibinfo{pages}{111} (\bibinfo{year}{1993}).

\bibitem[{\citenamefont{Epelbaum et~al.}(2005)\citenamefont{Epelbaum, Glockle,
  and Meissner}}]{Epelbaum:2004fk}
\bibinfo{author}{\bibfnamefont{E.}~\bibnamefont{Epelbaum}},
  \bibinfo{author}{\bibfnamefont{W.}~\bibnamefont{Glockle}}, \bibnamefont{and}
  \bibinfo{author}{\bibfnamefont{U.-G.} \bibnamefont{Meissner}},
  \bibinfo{journal}{Nucl. Phys.} \textbf{\bibinfo{volume}{A747}},
  \bibinfo{pages}{362} (\bibinfo{year}{2005}).

\bibitem[{\citenamefont{Entem and Machleidt}(2003)}]{Entem:2003ft}
\bibinfo{author}{\bibfnamefont{D.~R.} \bibnamefont{Entem}} \bibnamefont{and}
  \bibinfo{author}{\bibfnamefont{R.}~\bibnamefont{Machleidt}},
  \bibinfo{journal}{Phys. Rev.} \textbf{\bibinfo{volume}{C68}},
  \bibinfo{pages}{041001} (\bibinfo{year}{2003}).

\bibitem[{\citenamefont{Bogner et~al.}(2003)\citenamefont{Bogner, Kuo, and
  Schwenk}}]{Bogner:2003wn}
\bibinfo{author}{\bibfnamefont{S.~K.} \bibnamefont{Bogner}},
  \bibinfo{author}{\bibfnamefont{T.~T.~S.} \bibnamefont{Kuo}},
  \bibnamefont{and} \bibinfo{author}{\bibfnamefont{A.}~\bibnamefont{Schwenk}},
  \bibinfo{journal}{Phys. Rept.} \textbf{\bibinfo{volume}{386}},
  \bibinfo{pages}{1} (\bibinfo{year}{2003}).

\bibitem[{\citenamefont{Baru et~al.}(2009)\citenamefont{Baru, Epelbaum,
  Haidenbauer, Hanhart, Kudryavtsev, Lensky, and Meissner}}]{Baru:2009fm}
\bibinfo{author}{\bibfnamefont{V.}~\bibnamefont{Baru}},
  \bibinfo{author}{\bibfnamefont{E.}~\bibnamefont{Epelbaum}},
  \bibinfo{author}{\bibfnamefont{J.}~\bibnamefont{Haidenbauer}},
  \bibinfo{author}{\bibfnamefont{C.}~\bibnamefont{Hanhart}},
  \bibinfo{author}{\bibfnamefont{A.}~\bibnamefont{Kudryavtsev}},
  \bibinfo{author}{\bibfnamefont{V.}~\bibnamefont{Lensky}}, \bibnamefont{and}
  \bibinfo{author}{\bibfnamefont{U.}~\bibnamefont{Meissner}},
  \bibinfo{journal}{Phys.Rev.} \textbf{\bibinfo{volume}{C80}},
  \bibinfo{pages}{044003} (\bibinfo{year}{2009}), \eprint{0907.3911}.

\bibitem[{\citenamefont{Bolton and Miller}(2010{\natexlab{a}})}]{Bolton:2009rq}
\bibinfo{author}{\bibfnamefont{D.~R.} \bibnamefont{Bolton}} \bibnamefont{and}
  \bibinfo{author}{\bibfnamefont{G.~A.} \bibnamefont{Miller}},
  \bibinfo{journal}{Phys. Rev.} \textbf{\bibinfo{volume}{C81}},
  \bibinfo{pages}{014001} (\bibinfo{year}{2010}{\natexlab{a}}).

\bibitem[{\citenamefont{Filin et~al.}(2009)}]{Filin:2009yh}
\bibinfo{author}{\bibfnamefont{A.}~\bibnamefont{Filin}} \bibnamefont{et~al.},
  \bibinfo{journal}{Phys. Lett.} \textbf{\bibinfo{volume}{B681}},
  \bibinfo{pages}{423} (\bibinfo{year}{2009}).

\bibitem[{\citenamefont{Cohen et~al.}(1996)\citenamefont{Cohen, Friar, Miller,
  and van Kolck}}]{Cohen:1995cc}
\bibinfo{author}{\bibfnamefont{T.~D.} \bibnamefont{Cohen}},
  \bibinfo{author}{\bibfnamefont{J.~L.} \bibnamefont{Friar}},
  \bibinfo{author}{\bibfnamefont{G.~A.} \bibnamefont{Miller}},
  \bibnamefont{and} \bibinfo{author}{\bibfnamefont{U.}~\bibnamefont{van
  Kolck}}, \bibinfo{journal}{Phys. Rev.} \textbf{\bibinfo{volume}{C53}},
  \bibinfo{pages}{2661} (\bibinfo{year}{1996}).

\bibitem[{\citenamefont{Hanhart}(2004)}]{Hanhart:2003pg}
\bibinfo{author}{\bibfnamefont{C.}~\bibnamefont{Hanhart}},
  \bibinfo{journal}{Phys. Rept.} \textbf{\bibinfo{volume}{397}},
  \bibinfo{pages}{155} (\bibinfo{year}{2004}).

\bibitem[{\citenamefont{Koltun and Reitan}(1966)}]{Koltun:1965yk}
\bibinfo{author}{\bibfnamefont{D.~S.} \bibnamefont{Koltun}} \bibnamefont{and}
  \bibinfo{author}{\bibfnamefont{A.}~\bibnamefont{Reitan}},
  \bibinfo{journal}{Phys. Rev.} \textbf{\bibinfo{volume}{141}},
  \bibinfo{pages}{1413} (\bibinfo{year}{1966}).

\bibitem[{\citenamefont{Bolton and Miller}(2010{\natexlab{b}})}]{Bolton:2010qu}
\bibinfo{author}{\bibfnamefont{D.~R.} \bibnamefont{Bolton}} \bibnamefont{and}
  \bibinfo{author}{\bibfnamefont{G.~A.} \bibnamefont{Miller}},
  \bibinfo{journal}{Phys. Rev.} \textbf{\bibinfo{volume}{C82}},
  \bibinfo{pages}{024001} (\bibinfo{year}{2010}{\natexlab{b}}).

\bibitem[{\citenamefont{Gardestig et~al.}(2006)\citenamefont{Gardestig,
  Phillips, and Elster}}]{Gardestig:2005sn}
\bibinfo{author}{\bibfnamefont{A.}~\bibnamefont{Gardestig}},
  \bibinfo{author}{\bibfnamefont{D.~R.} \bibnamefont{Phillips}},
  \bibnamefont{and} \bibinfo{author}{\bibfnamefont{C.}~\bibnamefont{Elster}},
  \bibinfo{journal}{Phys. Rev.} \textbf{\bibinfo{volume}{C73}},
  \bibinfo{pages}{024002} (\bibinfo{year}{2006}).

\bibitem[{\citenamefont{Beane and Savage}(2003)}]{Beane:2002vs}
\bibinfo{author}{\bibfnamefont{S.~R.} \bibnamefont{Beane}} \bibnamefont{and}
  \bibinfo{author}{\bibfnamefont{M.~J.} \bibnamefont{Savage}},
  \bibinfo{journal}{Nucl. Phys.} \textbf{\bibinfo{volume}{A713}},
  \bibinfo{pages}{148} (\bibinfo{year}{2003}).

\bibitem[{\citenamefont{Bulgac et~al.}(1997)\citenamefont{Bulgac, Miller, and
  Strikman}}]{Bulgac:1997ji}
\bibinfo{author}{\bibfnamefont{A.}~\bibnamefont{Bulgac}},
  \bibinfo{author}{\bibfnamefont{G.~A.} \bibnamefont{Miller}},
  \bibnamefont{and} \bibinfo{author}{\bibfnamefont{M.}~\bibnamefont{Strikman}},
  \bibinfo{journal}{Phys.Rev.} \textbf{\bibinfo{volume}{C56}},
  \bibinfo{pages}{3307} (\bibinfo{year}{1997}).

\bibitem[{\citenamefont{Partovi and Lomon}(1970)}]{Partovi:1969wd}
\bibinfo{author}{\bibfnamefont{M.~H.} \bibnamefont{Partovi}} \bibnamefont{and}
  \bibinfo{author}{\bibfnamefont{E.~L.} \bibnamefont{Lomon}},
  \bibinfo{journal}{Phys. Rev.} \textbf{\bibinfo{volume}{D2}},
  \bibinfo{pages}{1999} (\bibinfo{year}{1970}).

\bibitem[{\citenamefont{Miller and Tiburzi}(2010)}]{Miller:2009fc}
\bibinfo{author}{\bibfnamefont{G.~A.} \bibnamefont{Miller}} \bibnamefont{and}
  \bibinfo{author}{\bibfnamefont{B.~C.} \bibnamefont{Tiburzi}},
  \bibinfo{journal}{Phys. Rev.} \textbf{\bibinfo{volume}{C81}},
  \bibinfo{pages}{035201} (\bibinfo{year}{2010}).

\bibitem[{\citenamefont{Jenkins and Manohar}(1991)}]{Jenkins:1990jv}
\bibinfo{author}{\bibfnamefont{E.~E.} \bibnamefont{Jenkins}} \bibnamefont{and}
  \bibinfo{author}{\bibfnamefont{A.~V.} \bibnamefont{Manohar}},
  \bibinfo{journal}{Phys. Lett.} \textbf{\bibinfo{volume}{B255}},
  \bibinfo{pages}{558} (\bibinfo{year}{1991}).

\bibitem[{\citenamefont{Scherer}(2003)}]{Scherer:2002tk}
\bibinfo{author}{\bibfnamefont{S.}~\bibnamefont{Scherer}},
  \bibinfo{journal}{Adv. Nucl. Phys.} \textbf{\bibinfo{volume}{27}},
  \bibinfo{pages}{277} (\bibinfo{year}{2003}).

\bibitem[{\citenamefont{Horowitz}(1993)}]{Horowitz:1993sh}
\bibinfo{author}{\bibfnamefont{C.}~\bibnamefont{Horowitz}},
  \bibinfo{journal}{Phys.Rev.} \textbf{\bibinfo{volume}{C48}},
  \bibinfo{pages}{2920} (\bibinfo{year}{1993}).

\bibitem[{\citenamefont{Lee and Riska}(1993)}]{Lee:1993xh}
\bibinfo{author}{\bibfnamefont{T.~S.~H.} \bibnamefont{Lee}} \bibnamefont{and}
  \bibinfo{author}{\bibfnamefont{D.~O.} \bibnamefont{Riska}},
  \bibinfo{journal}{Phys. Rev. Lett.} \textbf{\bibinfo{volume}{70}},
  \bibinfo{pages}{2237} (\bibinfo{year}{1993}).

\bibitem[{\citenamefont{Wiringa et~al.}(1995)\citenamefont{Wiringa, Stoks, and
  Schiavilla}}]{Wiringa:1994wb}
\bibinfo{author}{\bibfnamefont{R.~B.} \bibnamefont{Wiringa}},
  \bibinfo{author}{\bibfnamefont{V.~G.~J.} \bibnamefont{Stoks}},
  \bibnamefont{and}
  \bibinfo{author}{\bibfnamefont{R.}~\bibnamefont{Schiavilla}},
  \bibinfo{journal}{Phys. Rev.} \textbf{\bibinfo{volume}{C51}},
  \bibinfo{pages}{38} (\bibinfo{year}{1995}).

\bibitem[{\citenamefont{Lensky et~al.}(2006)}]{Lensky:2005jc}
\bibinfo{author}{\bibfnamefont{V.}~\bibnamefont{Lensky}} \bibnamefont{et~al.},
  \bibinfo{journal}{Eur. Phys. J.} \textbf{\bibinfo{volume}{A27}},
  \bibinfo{pages}{37} (\bibinfo{year}{2006}).

\bibitem[{\citenamefont{Hanhart and Baru}(2010)}]{Hanhart}
\bibinfo{author}{\bibfnamefont{C.}~\bibnamefont{Hanhart}} \bibnamefont{and}
  \bibinfo{author}{\bibfnamefont{V.}~\bibnamefont{Baru}},
  \bibinfo{howpublished}{private communications} (\bibinfo{year}{2010}).

\bibitem[{\citenamefont{Friar et~al.}(1984)\citenamefont{Friar, Gibson, and
  Payne}}]{Friar:1984wi}
\bibinfo{author}{\bibfnamefont{J.~L.} \bibnamefont{Friar}},
  \bibinfo{author}{\bibfnamefont{B.~F.} \bibnamefont{Gibson}},
  \bibnamefont{and} \bibinfo{author}{\bibfnamefont{G.~L.} \bibnamefont{Payne}},
  \bibinfo{journal}{Phys. Rev.} \textbf{\bibinfo{volume}{C30}},
  \bibinfo{pages}{1084} (\bibinfo{year}{1984}).

\bibitem[{\citenamefont{Hanhart et~al.}(2001)\citenamefont{Hanhart, Miller,
  Myhrer, Sato, and van Kolck}}]{Hanhart:2000wf}
\bibinfo{author}{\bibfnamefont{C.}~\bibnamefont{Hanhart}},
  \bibinfo{author}{\bibfnamefont{G.~A.} \bibnamefont{Miller}},
  \bibinfo{author}{\bibfnamefont{F.}~\bibnamefont{Myhrer}},
  \bibinfo{author}{\bibfnamefont{T.}~\bibnamefont{Sato}}, \bibnamefont{and}
  \bibinfo{author}{\bibfnamefont{U.}~\bibnamefont{van Kolck}},
  \bibinfo{journal}{Phys. Rev.} \textbf{\bibinfo{volume}{C63}},
  \bibinfo{pages}{044002} (\bibinfo{year}{2001}).

\bibitem[{\citenamefont{Stoks et~al.}(1994)\citenamefont{Stoks, Klomp,
  Terheggen, and de~Swart}}]{Stoks:1994wp}
\bibinfo{author}{\bibfnamefont{V.~G.~J.} \bibnamefont{Stoks}},
  \bibinfo{author}{\bibfnamefont{R.~A.~M.} \bibnamefont{Klomp}},
  \bibinfo{author}{\bibfnamefont{C.~P.~F.} \bibnamefont{Terheggen}},
  \bibnamefont{and} \bibinfo{author}{\bibfnamefont{J.~J.}
  \bibnamefont{de~Swart}}, \bibinfo{journal}{Phys. Rev.}
  \textbf{\bibinfo{volume}{C49}}, \bibinfo{pages}{2950} (\bibinfo{year}{1994}).

\bibitem[{\citenamefont{Hutcheon et~al.}(1990)}]{Hutcheon:1989bt}
\bibinfo{author}{\bibfnamefont{D.~A.} \bibnamefont{Hutcheon}}
  \bibnamefont{et~al.}, \bibinfo{journal}{Phys. Rev. Lett.}
  \textbf{\bibinfo{volume}{64}}, \bibinfo{pages}{176} (\bibinfo{year}{1990}).

\bibitem[{\citenamefont{Heimberg et~al.}(1996)}]{Heimberg:1996be}
\bibinfo{author}{\bibfnamefont{P.}~\bibnamefont{Heimberg}}
  \bibnamefont{et~al.}, \bibinfo{journal}{Phys. Rev. Lett.}
  \textbf{\bibinfo{volume}{77}}, \bibinfo{pages}{1012} (\bibinfo{year}{1996}).

\bibitem[{\citenamefont{Drochner et~al.}(1998)}]{Drochner:1998ja}
\bibinfo{author}{\bibfnamefont{M.}~\bibnamefont{Drochner}} \bibnamefont{et~al.}
  (\bibinfo{collaboration}{GEM}), \bibinfo{journal}{Nucl. Phys.}
  \textbf{\bibinfo{volume}{A643}}, \bibinfo{pages}{55} (\bibinfo{year}{1998}).

\bibitem[{\citenamefont{Strauch et~al.}(2010)}]{Strauch:2010rm}
\bibinfo{author}{\bibfnamefont{T.}~\bibnamefont{Strauch}} \bibnamefont{et~al.},
  \bibinfo{journal}{Phys. Rev. Lett.} \textbf{\bibinfo{volume}{104}},
  \bibinfo{pages}{142503} (\bibinfo{year}{2010}).

\bibitem[{\citenamefont{Ericson and Weise}(1988)}]{Ericson:1988gk}
\bibinfo{author}{\bibfnamefont{T.~E.~O.} \bibnamefont{Ericson}}
  \bibnamefont{and} \bibinfo{author}{\bibfnamefont{W.}~\bibnamefont{Weise}},
  \emph{\bibinfo{title}{Pions and Nuclei}}, vol.~\bibinfo{volume}{74} of
  \emph{\bibinfo{series}{The International Series of Monographs on Physics}}
  (\bibinfo{publisher}{Clarendon}, \bibinfo{address}{Oxford},
  \bibinfo{year}{1988}).

\end{thebibliography}

\end{document}